\documentclass[12pt,sort,compress]{elsarticle}

\usepackage{graphicx}
\usepackage{amsmath,amssymb}

\usepackage{enumitem}
\usepackage{mdframed}

\newcommand{\ud}{\mathrm{d}}

\definecolor{grey}{rgb}{.9,.9,.9}
\definecolor{darkgrey}{rgb}{.3,.3,.3}

\newmdenv[linecolor=black, backgroundcolor = grey, linewidth=1pt, innerbottommargin = 10pt, leftmargin = -2.5cm, rightmargin = -2.5cm, nobreak]{infobox}

\journal{Physics of Life Reviews}

\begin{document}

\begin{frontmatter}


\title{From genotypes to organisms: State-of-the-art and perspectives of a cornerstone in evolutionary dynamics}



\author[a,b]{Susanna Manrubia\fnref{1,2}}
\author[b,c,d,e]{Jos\'e A. Cuesta\fnref{1}}

\fntext[1]{Organizing and leading authors; contributed equally to this work}
\fntext[2]{Corresponding author; smanrubia@cnb.csic.es}

\author[b,f]{Jacobo Aguirre}
\author[g,h]{Sebastian E. Ahnert}
\author[i]{Lee Altenberg}
\author[j,k]{Alejandro V. Cano}
\author[b,c]{Pablo Catal\'an}
\author[l,m]{Ramon Diaz-Uriarte}
\author[n,o]{Santiago F. Elena}
\author[p]{Juan Antonio Garc\'{\i}a-Mart\'{\i}n}
\author[q]{Paulien Hogeweg}
\author[r,s]{Bhavin S. Khatri}
\author[t]{Joachim Krug}
\author[u]{Ard A. Louis}
\author[g,h]{Nora S. Martin}
\author[j,k]{Joshua L. Payne}
\author[v]{Matthew J. Tarnowski}
\author[g,h]{Marcel Wei{\ss}}

\address[a]{Department of Systems Biology, Centro Nacional de Biotecnolog\'{\i}a (CSIC), Madrid, Spain}
\address[b]{Grupo Interdisciplinar de Sistemas Complejos (GISC), Madrid, Spain}
\address[c]{Departamento de Matem\'aticas, Universidad Carlos III de Madrid, Legan\'es, Spain}
\address[d]{Instituto de Biocomputaci\'on y F\'{\i}sica de Sistemas Complejos (BiFi), Universidad de Zaragoza, Spain}
\address[e]{UC3M-Santander Big Data Institute (IBiDat), Getafe, Madrid, Spain}
\address[f]{Centro de Astrobiología, CSIC-INTA, ctra. de Ajalvir km 4, 28850 Torrejón de Ardoz, Madrid, Spain}
\address[g]{Theory of Condensed Matter Group, Cavendish Laboratory, University of Cambridge, Cambridge, UK}
\address[h]{Sainsbury Laboratory, University of Cambridge, Cambridge, UK}
\address[i]{University of Hawai`i at Manoa, Hawai, US}
\address[j]{Institute of Integrative Biology, ETH Zurich, Zurich, Switzerland}
\address[k]{Swiss Institute of Bioinformatics, Lausanne, Switzerland}
\address[l]{Department of Biochemistry, Universidad Aut\'onoma de Madrid, Madrid, Spain}
\address[m]{Instituto de Investigaciones Biom\'edicas ``Alberto Sols" (UAM-CSIC), Madrid, Spain}
\address[n]{Instituto de Biolog\'{\i}a Integrativa de Sistemas, I$^2$SysBio (CSIC-UV), Val\`encia, Spain}
\address[o]{The Santa Fe Institute, Santa Fe, New Mexico, USA}
\address[p]{Centro Nacional de Biotecnolog\'{\i}a (CSIC), Madrid, Spain}
\address[q]{Theoretical Biology and Bioinformatics Group, Utrecht University, The Netherlands}
\address[r]{The Francis Crick Institute, London, UK}
\address[s]{Department of Life Sciences, Imperial College London, London, UK}
\address[t]{Institute for Biological Physics, University of Cologne, K\"oln, Germany}
\address[u]{Rudolf Peierls Centre for Theoretical Physics, University of Oxford, Oxford, UK}
\address[v]{School of Biological Sciences, University of Bristol, Bristol, UK}

\begin{abstract}
Understanding how genotypes map onto phenotypes, fitness, and eventually organisms is arguably the next major missing piece in a fully predictive theory of evolution. We refer to this generally as the problem of the genotype-phenotype map.  Though we are still far from achieving a complete picture of these relationships, our current understanding of simpler questions, such as the structure induced in the space of genotypes by sequences mapped to molecular structures, has revealed important facts that deeply affect the dynamical description of evolutionary processes. Empirical evidence supporting the fundamental relevance of features such as phenotypic bias is mounting as well, while the synthesis of conceptual and experimental progress leads to questioning current assumptions on the nature of evolutionary dynamics---cancer progression models or synthetic biology approaches being notable examples. This work delves into a critical and constructive attitude in our current knowledge of how genotypes map onto molecular phenotypes and organismal functions, and discusses theoretical and empirical avenues to broaden and improve this comprehension. As a final goal, this community should aim at deriving an updated picture of evolutionary processes soundly relying on the structural properties of genotype spaces, as revealed by modern techniques of molecular and functional analysis.
\end{abstract}


\end{frontmatter}


\section{Introduction}
\label{S:1}

How genetic variation contributes to phenotypic variation is an essential question that must be answered to understand the evolutionary process. The experimental characterisation of the genotype-phenotype (GP) relationship is a formidable theoretical and experimental challenge, but also an expensive task which suffers from severe practical limitations. Computational approaches have been recurrently used to make predictions of phenotypes from genotypes and to uncover the statistical features of that relationship. Advances notwithstanding, an apparently insurmountable problem remains: the astronomically large size of the space of genotypes. The space of possible phenotypic change and the probabilities of such change are directly determined by the architecture of the GP map; to quantify this map will allow better quantification of how the space of phenotypes is explored and answer important questions about the probability of evolutionary rescue or innovation under endogenous or exogenous changes.

Progress in our understanding of GP maps at various levels is of relevance for different scientific communities with interests that range from evolutionary theory to molecular design through genomic bases of disease aetiology. 
Understanding of how RNA, DNA or amino acid sequences map onto molecular function could be of great importance for more fundamental approaches in synthetic biology, biotechnology, and systems chemistry. In a broader ecological context, the way in which generic properties of the GP map shape adaptation have rarely been explored. As of today, the overarching question of whether organismal phenotypes can be predicted from microscopic properties of genotype spaces remains open.

In this review, we discuss the state-of-the-art of genotype-to-organism research and future research avenues in the field. The review is structured into four major parts. The first part is constituted by this introduction and Section~\ref{sec:variation}, which puts in perspective how relevant the generation of variation is in the evolutionary process, and introduces important biases arising from the inherent structure of genotype spaces. 

The second part comprises sections~\ref{sec:models} to \ref{sec:evolutionOFgpmaps}, where we discuss conceptual approaches to the static properties of GP maps and their dynamical consequences, as well as the evolution of GP maps themselves. The field is broad and several aspects have been addressed in previous reviews, so we only briefly summarise topics dealt with elsewhere. Therefore, we will succinctly present computational GP maps and only recapitulate, taking an integrative and explanatory viewpoint, the topological properties of the space of genotypes \cite{reidys:1997,stadler:2006,wagner:2011,ahnert:2017,aguirre:2018,nichol:2019}. Section~\ref{sec:models} constitutes a synthetic overview of GP map models, including paradigmatic examples such as RNA folding, more recent multi-level models such as toyLIFE, and a summary of artificial life examples. Readers familiar with those models can safely skip that section. Those models endow genotype spaces with topological properties that are briefly reviewed in the introduction of Section~\ref{sec:UnivTopology}, which is mostly devoted to discussing possible roots for generic properties of a broad class of GP maps. Attention is subsequently devoted to population dynamics on genotype spaces, which has been a less explored topic. Section~\ref{sec:dynamics} describes transient and equilibrium dynamical features of evolutionary processes. First, it delves into the effects of recombination and mutation bias, and on phenotypic transitions caused by the hierarchical, networked structure of genotype spaces. Then, a mean-field description that incorporates the essentials of GP map topology to clarify major dynamical features is discussed. The section finishes with a derivation of equilibrium properties in the context of statistical mechanics and some applied examples. Section~\ref{sec:evolutionOFgpmaps} discusses the evolution of GP maps themselves by means of two illustrative examples: a scenario where a multifunctional quasispecies emerges and a model of virtual cells incorporating the evolution of genome size.

The third part, sections \ref{sec:empirical} and \ref{sec:cancer} is devoted to empirical GP maps and to biological applications, and mostly presents topics under development.  Section~\ref{sec:empirical} examines most recent achievements regarding the experimental characterisation of GP and genotype-to-function maps in molecules and simple organisms, and the different possibilities that current and future techniques might allow. It includes a formal discussion on how phenotypes can be inferred from genotypic data and fitness assays, and a discussion of the intimate relationship between fitness landscapes and GP maps. Section~\ref{sec:cancer} exemplifies how concepts and techniques originating in quantitative studies of the GP map can enlighten useful approaches to diseases with a genetic component. 

The fourth and last part presents a mostly self-contained overview of open questions and difficulties that the field faces, as well as some possible avenues for further progress, in
Section~\ref{sec:perspectives}. The paper closes with an outlook in Section~\ref{sec:GOmap} where we reflect on the feasibility of characterising the genotype-to-organism map, and on plausible epistemological difficulties to comprehend the organisation and complexity of full organisms. 

\section{GP maps and the importance of variation}
\label{sec:variation}

Darwinian evolution requires heritable phenotypic variation, upon which natural selection acts. Much of traditional evolutionary theory has focused on the role of natural selection, while the study  of variation has been much less developed. There are a number of reasons for this difference.

Firstly, there is an influential tradition, stemming from the early days of the modern synthesis, that any meaningful change over evolutionary time is ultimately caused by natural selection. One argument in favour of this thesis comes from the simple observation that a heritable phenotype with higher fitness will, over the generations, exponentially out-compete other phenotypes with lower fitness in the same population. Thus, differences in the rate at which mutations arrive will be swamped by the effect of fitness differences (there are much more sophisticated versions of this argument). Another argument, which is often more implicitly than explicitly made, is that a large part of variation is \textit{isotropic}---in other words, it is not biased in one direction or another. Stephen J. Gould, who was critical of this viewpoint, expresses it as follows: ``\textit{variation becomes raw material only, an isotropic sphere of potential about the modal form of a species \ldots [only] natural selection \ldots can manufacture substantial, directional change}'' \cite{gould:2002}. Whether evolutionary trends must primarily be explained by natural selection, or whether anisotropic (biased) variation also plays a key role, is a complex question. While the arguments have moved on considerably since the critique of Gould, especially with the rise of evo-devo \cite{love:2015}, they are far from being settled \cite{laland:2014,stoltzfus:2018}. Ever since the modern synthesis, directed variation has been deemed anathema because it evokes the Lamarckian view of variation to facilitate adaptation. However, as the analysis of GP maps reveals, these maps are a major source of anisotropic variation, even if this variation is not necessarily biased in the most beneficial way for the organism.

The second reason why our understanding of variation is relatively underdeveloped is that working out the exact role played by the arrival of variation in evolutionary history is difficult because in nature we typically only observe the final outcomes of an evolutionary process. It is hard to know what variation may have arisen in the past, but not fixed, or what variation could have potentially arisen, but did not. For example, even when all potential variation is isotropic, the non-lethal variation may well be anisotropic, depending on the environment.
 
In this context, the study of GP maps is critical, because they provide access to the way that changes in genotypes, brought on by various kinds of mutations, are translated into phenotypic variation for the biological system that the map describes. They allow us to ask  important \textit{counterfactual} questions, such as what is the full spectrum of variation that could potentially arise? Working out how variation affects evolutionary outcomes depends on an understanding of such counterfactuals.    

A final issue for understanding variation comes from the unfathomable vastness of genotype spaces, whose size grows exponentially with genome length, rapidly leading to hyperastronomical numbers of possibilities \cite{louis:2016}. If these spaces are so unimaginably vast, then it might seem natural to conclude, as many have done, that the variation that appears in evolutionary history is largely contingent upon accidents of history, and unlikely to be repeated (see Ref. \cite{louis:2016} for a discussion).

This problem of hyperastronomically large spaces means that only relatively simple GP maps allow global questions about the full spectrum of possible variation to be addressed. Nevertheless, important progress in this direction has been made through the use of GP maps that can be computationally explored and, more recently, through the development of quantitative approaches to shared generic properties. Among the latter, one of the most striking properties is a strong \textit{bias} in the number of genotypes mapping to a phenotype \cite{greenbury:2016,ahnert:2017}. This begs the question: Can this bias, which often extends over many orders of magnitude, affect evolutionary outcomes? Indeed, phenotypic bias, among other non-trivial properties of GP maps, does severely affect not only our understanding of how variation arises through random mutations, but also any accurate representation---be it metaphorical or formal---of evolutionary dynamics at large. 

\section{Models of the GP map}
\label{sec:models}

Maynard Smith introduced the notion of a mapping from a genetic space to a molecular structure---and with it the idea of a network linking viable genotypes---as a resolution of an evolutionary paradox pointed out by Salisbury \cite{maynard-smith:1970}. In brief, Salisbury noted \cite{salisbury:1969} that the number of possible amino acid sequences exceeds by many orders of magnitude the number of proteins that ever existed on Earth since the origin of life, and concluded from this fact that functionally effective proteins have a vanishingly small chance of arising by mutation. As a way out of this dilemma, Maynard Smith suggested that the existence of networks of functional proteins are essential to navigate the space of genotypes to produce a sequence of adaptive improvements and to explore new regions that, eventually, secure evolutionary innovation \cite{huynen:1996b}. Formally, the space of genotypes can be defined as a network where nodes represent genotypes, with any two nodes linked if they are mutually accessible through a single point mutation \cite{schuster:1994}. A \emph{neutral network} is therefore an ensemble of connected genotypes with the same fitness, including those with identical phenotypes. The empirical existence of such networks and their role in providing access to new phenotypes \cite{fontana:1998} was unequivocally demonstrated \cite{koelle:2006,schultes:2000} four decades after Maynard Smith's conjecture.

Many studies have aimed at probing the statistical structure of the GP relationship, thus relying on the computational exploration of GP maps. Models of RNA secondary structure \cite{hofacker:1994,schuster:1994}, protein secondary structures \cite{lipman:1991,irback:2002}, gene regulatory networks \cite{wagner:2011,payne:2014a}, metabolic networks \cite{barve:2013,hosseini:2015}, protein complexes \cite{johnston:2011,greenbury:2014}, artificial life \cite{ofria:2004}, or multilevel maps such as a toyLIFE, which includes protein structure, regulatory, and metabolic networks \cite{arias:2014,catalan:2018}, have been explored through the years. Computational frameworks often rely on building complete GP maps from exhaustive enumeration of genotypes (or sparse GP maps from large samples) in models with simple genotype-to-phenotype rules as the ones above. To study global properties of a GP map, such as phenotype frequencies, a large number of genotype-phenotype pairs have to be evaluated. With notable exceptions \cite{aguilar:2017,rowe:2009,jimenez:2013}, some of which will be discussed in section~\ref{sec:empirical} of this paper, the exhaustive study of GP maps represents an enormous challenge that has been restricted to systems where the phenotype can be found computationally from the genotypic information. 

For the sake of simplicity, most GP computational maps assign a unique phenotype to each genotype, in a many-to-one representation. Some maps also take into account environmental factors such as temperature, which modify GP mapping rules and, therefore, include phenotypic plasticity in a streamlined fashion \cite{wagner:2014}. Other implementations also consider phenotypic promiscuity \cite{jensen:1976,aharoni:2005}, that is, the possibility that each sequence maps to more than one phenotype under fixed environmental variables. However, many-to-many GP maps entail an exponentially increasing cost in computation time, so they have been rarely explored in depth (for exceptions see \cite{ancel:2000,barve:2013,wagner:2014,espinosa-soto:2011,deboer:2012,deboer:2014,payne:2014c,rezazadegan:2018,diaz-uriarte:2018}).

Creating complete computational frameworks for GP models is a challenge---building complete GP maps for sequences as long as functional molecules in realistic environments is beyond our current computational power. Nevertheless, progress has been steady and significant. For example, and despite the freedom inherent to any definition of phenotype, many generalities have emerged from studying these models, and theoretical arguments to explain some of them have been developed \cite{greenbury:2015,manrubia:2017,garcia-martin:2018}. These studies have led to a relatively sound understanding of the conditions that are behind different phenotype abundances, its relationship with robustness, and the topology of neutral networks \cite{greenbury:2016,aguirre:2009,aguirre:2011}.

In this section, we begin by briefly summarising a variety of GP maps that have been computationally studied to date. Some attention is devoted to RNA, a model for which we examine in perspective some of the important lessons learnt and discuss possible future contributions to GP map research. There is a substantial body of literature available, including comprehensive reviews \cite{schuster:2006} that we do not even attempt to summarise here. We finish this part discussing the GP maps of artificial life systems.

We would like to remark that, while we focus here on sequence-to-structure and sequence-to-function maps, it is important to highlight that genotype-phe\-no\-type and genotype-fitness maps have also been studied in the context of development \cite{salazar-ciudad:2010,salazar-ciudad:2013,hagolani:2021}.

\subsection{One-level GP models}

\begin{figure}[ht]
\begin{center}
\includegraphics[width=7.0cm]{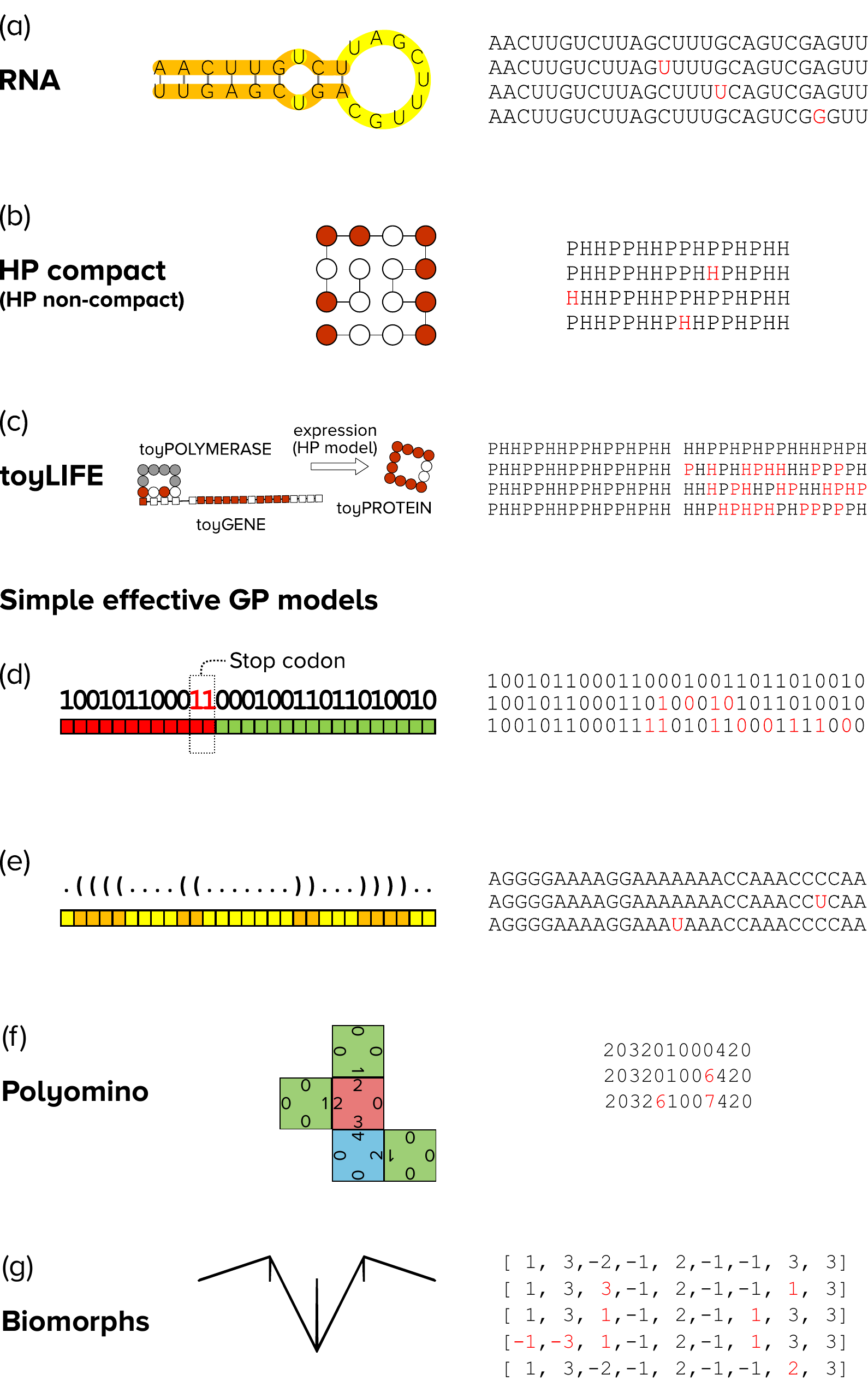}
\end{center}
\caption{Some examples of simple GP maps. For each model, and from left to right, we depict an example phenotype and some of the genotypes in its neutral network (mutations that do not change the phenotype are highlighted in red). (a) RNA sequence-to-structure is the paradigmatic GP map. Mutations that conserve the secondary structure appear in loops with a higher likelihood than in stems. (b) The HP model, both in compact or non-compact realisations, has been studied as a model for protein folding. (c) toyLIFE is a minimal model with several levels \cite{arias:2014}. Sequences of the HP type are read and translated to proteins that interact through analogous HP rules to break down metabolites. (d) Fibonacci's model \cite{greenbury:2015} relies on the separation between constrained and unconstrained positions in sequences to derive some formal properties of simple GP maps. (e) A generalisation of the idea of position-dependent constraints \cite{manrubia:2017} provides a formal understanding of the ubiquitous lognormal distribution for neutral set sizes. (f) A polyomino model used to capture the essentials of quaternary protein structure \cite{greenbury:2014}. (g) Dawkins' biomorphs are defined by genotypes with few parameters that define the generative rules of the structure \cite{dawkins:2003}. Figure modified from Ref. \cite{aguirre:2018}.}
\label{fig:GPmodels}
\end{figure}

Over the past three decades, the GP maps of several simple biological model systems have been studied in great detail. Figure~\ref{fig:GPmodels} summarises the essentials of some of the GP maps we will be discussing. Two classical examples are RNA secondary structure \cite{hofacker:1994,schuster:1994} and the HP model of protein folding \cite{lipman:1991,li:1996}. The HP model represents proteins on a regular lattice as self-avoiding chains of hydrophobic (H) or polar (P) beads. In its compact version the chains are forced to fold into rectangular configurations that leave no empty sites, while in the non-compact version all possible self-avoiding walks in the lattice are considered. The phenotype is defined as the minimum energy of a given configuration calculated from a contact potential between neighbouring (but not in the backbone) beads. Because RNA and HP models are relatively tractable, properties such as the distribution of the number of genotypes per phenotype \cite{stich:2008,louis:2016,shahrezaei:1999}, the phenotypic robustness and evolvability \cite{wagner:2007,holzgrafe:2011} (see Box~\ref{box:definitions}) or the topological structure of neutral networks \cite{aguirre:2011} could be systematically studied and compared \cite{ferrada:2012}.

Given the pivotal role proteins play in cellular processes, the protein sequence-to-structure map, of which the HP model constitutes the simplest realisation, is of great general interest \cite{shakhnovich:2006,ciliberti:2007b,chen:2008}. The protein sequence-to-structure map has been also studied using more realistic, multi-parametric contact potentials \cite{mirny:1996,buchler:1999,li:2002,bastolla:2003} and coarse-grained models at different levels, such as the Polyomino model \cite{johnston:2011,greenbury:2014} for protein complexes. 
Some inferences about local and global properties of the protein sequence-to-structure-to-function GP map have also been made from experimental data \cite{sarkisyan:2016,ferrada:2008,ferrada:2010}, and estimates of neutral set sizes (NSSs) have been obtained from structural data \cite{england:2003}.

Breakthroughs in the computational prediction of protein structure from amino acid sequence have recently been achieve by deep learning with artificial neural networks. The AlphaFold 2 system by DeepMind \cite{callaway:2020} outperformed 100 other teams in the 2020 Critical Assessment of Structure Prediction challenge (CASP13), with prediction accuracy rivaling that of experimental structure determination. Enormous resources are required for each evaluation, however, taking days of real time computational for a single protein sequence. If subsequent development can maintain the accuracy while allowing exponential speedups, these computational systems should be able to open up entirely new investigations of the GP map for proteins.

\begin{figure}
\begin{infobox}[frametitle= Box~\ref{box:definitions}a. Definitions]
\begin{description}[leftmargin=5mm]\setlength{\itemsep}{0pt}
\item[\it Function] Function is a contentious term \cite{graur:2013,kellis:2014} that is used to mean many things. In this review we are mostly referring to properties of proteins, such as stability, catalytic activity, and binding affinity.
\item[\it Genetic correlations] A GP map has this property if two sequences differing at a single site are more likely to generate the same phenotype than two arbitrary sequences \cite{greenbury:2016}.
\item[\it Genotype network] A set of mutually connected genotypes that have the same phenotype. This term is usually employed as a synonym of \emph{neutral network,} although in some context a genotype network needs not be neutral---for instance, in the case of GP maps with both a categorical phenotype (e.g. molecular structure) as well as a quantitative fitness (e.g. thermodynamic stability of the structure).
\item[\it Genotypic evolvability] Total number of distinct alternative phenotypes that can be reached through point mutations from a single genotype \cite{wagner:2007}.
\item[\it Genotypic robustness] Number of point mutations that do not change the phenotype of a specific genotype. It is analogous to the neutrality of a genotype.
\item[\it Navigability] Ability to navigate throughout genotype space via neutral mutations.
\item[\it Neutral network] A set of mutationally connected genotypes that have the same fitness, including those that have the same phenotype. Often, it refers to the largest connected component of a neutral set.
\item[\it Neutral set] A set of genotypes which have the same fitness, including those that have the same phenotype. The \textit{neutral set size} is therefore the number of genotypes that map to a given phenotype.
\item[\it Organism] Any individual entity that embodies the properties of life, like a cell, an animal, or a plant. It is a synonym for ``life form''. By extension, it also applies to artificial life forms.
\item[\it Phenotype] A property which is encoded in the genotype and is biologically relevant, for example a molecular structure. Though abstract, this broad definition allows a variety of models to be treated with the same terminology.
\item[\it Phenotypic robustness] Average genotypic robustness of all genotypes in a neutral network \cite{wagner:2007}.
\item[\it Phenotypic evolvability] Total number of distinct alternative phenotypes that can be reached through point mutations from a phenotype's neutral network \cite{wagner:2007}.
\item[\it Plasticity] Quality of a genotype leading to the production of more than one phenotype depending on the environment \cite{rezazadegan:2018}.
\item[\it Promiscuity] Quality of a genotype leading to the production of more than one phenotype in the same environment.
\item[\it Quasispecies] Population structure with a large numbers of variant genomes related by mutations. Quasispecies typically arise under high mutation rates as possible mutants change in relative frequency as replication and selection proceeds \cite{domingo:2019}.
\item[\it Shape-space-covering] A GP map has the shape space covering property if, given a phenotype, only a small radius around a sequence encoding that phenotype needs to be explored in order to find the most common phenotypes \cite{schuster:1994}.
\item[\it Versatility] A quantitative measure of the rescaled robustness of a specific sequence position \cite{garcia-martin:2018}.
\end{description}
\label{box:definitions}
\end{infobox}
\end{figure}
\begin{figure}
\begin{infobox}[frametitle= Box~\ref{box:definitions}b. Acronyms]
\begin{description}\setlength{\itemsep}{0pt}
\item[\rm CPMs] Cancer progression models
\item[\rm DAG] Directed acyclic graph
\item[\rm FACS] Fluorescence-activated cell sorting
\item[\rm FPGA] Field-programmable gate array
\item[\rm GP] Genotype-to-phenotype
\item[\rm MAVEs] Multiplexed assays for variant effects
\item[\rm MFE] Minimum free energy
\item[\rm MPRAs] Massively parallel reporter assays
\item[\rm NSS] Neutral set size
\item[\rm OLS] Oligo(nucleotide) library synthesis
\item[\rm SCRaMbLE] Synthetic Chromosome Recombination and Modification by LoxP-mediated Evolution
\end{description}
\end{infobox}
\end{figure}

A number of models work at levels above sequences. Simple gene regulatory networks act as effective genotypes in models that map them onto phenotypes defined as the steady-state gene expression pattern \cite{wagner:2003,ciliberti:2007}. A metabolic genotype is defined as all chemical reactions an organism can catalyse via enzymes encoded in its genome; the phenotype is defined as viability in minimal chemical environments that differ in their sole carbon sources \cite{matias-rodrigues:2009,samal:2010}. Those two models share the property that most genotypes do not map to any functional phenotype---it has been put forward that such a restrictive relationship may stem from a minimisation of the cost incurred by maintaining a complex functional network \cite{leclerc:2008}. However, genotype spaces where function is sparse still contain large neutral networks that percolate that space and guarantee phenotypic innovation without loss of function \cite{ciliberti:2007,matias-rodrigues:2009,barve:2013}.

There are compact \cite{catalan:2018} and non-compact \cite{holzgrafe:2011} versions of the HP model with an overwhelming majority of non-functional genotypes where neutral networks are very small and mostly disconnected; therefore, innovation is severely hindered, if not plainly impossible, in those one-level maps. However, that lack of navigability turns out to be irrelevant if additional, higher levels, are taken into account.

\subsection{Multi-level GP models}
\label{sec:multilevel}
Most computational GP maps studied to date, including those discussed in the previous section, only include one level (or scale) of description, mapping genotypes of different kinds to their corresponding phenotypes (see, however, \cite{serohijos:2014}). But even the simplest organisms include more than one level: RNAs and proteins will perform enzymatic and regulatory reactions that will in turn affect the availability of other molecules inside and outside the cell. If the study of one-level GP maps has led to great changes in our understanding of evolutionary theory, it stands to reason that studying multi-level GP maps will yield equally important insights.

It has been shown that multilevel models endowed with biophysically realistic interaction rules lead to the emergence of complex fitness landscapes that permit multiple, equally successful, evolutionary pathways \cite{heo:2008,heo:2011} or the growth of organismal population size when protein-based, functional genotypes are discovered through evolution \cite{zeldovich:2007_PLoSCB}. Recent proposals for multilevel models are the model of RNA-based virtual cells discussed in Section~\ref{sec:VirtualCells}, a model of developmental spatial patterning  \cite{khatri:2009,khatri:2019} (see Section~\ref{sec:SMevol}), and toyLIFE  \cite{arias:2014,catalan:2018}. toyLIFE is a multi-level model that includes genes, proteins and metabolites, as well as their regulatory and metabolic interactions. toyGenes consist of binary sequences (the genotype) that are first mapped to HP-like proteins. None of these proteins can be obtained from any other through single-point mutations. Proteins interact between themselves, with the genome, and with metabolites. The phenotype is defined by the set of metabolites that a given sequence is able to catabolise. In its three-gene version, phenotype is mostly defined through the first two genes, which admit very few mutations, while the third gene is essentially free to mutate, thus restoring evolvability to the system. Additionally, the existence of promiscuous sequences further enhances navigability when environmental factors such as temperature are considered \cite{catalan:2017tesis}. Promiscuity was recognised long ago as a key property in adaptive processes \cite{jensen:1976} that, as of yet, has not been explored in most GP maps. 

One of the most interesting results to come out of an early exploration of toyLIFE's metabolic GP map is that adding levels of complexity to a phenotypic definition actually increases robustness \cite{catalan:2018}: proteins can change and become non-functional, and regulatory functions can be altered, while the overall metabolic function remains constant. This suggests that the potential for cells to evolve toward new evolutionary challenges has been significantly underestimated in the past. 

\subsection{RNA}
\label{subsec:RNA}

RNA is the most paradigmatic model for studying GP relationships and constructing GP maps \cite{schuster:1994,fontana:1993,schultes:2005,wagner:2005,smit:2006,cowperthwaite:2008,jorg:2008,stich:2011,aguirre:2011,schaper:2014,dingle:2015,garcia-martin:2018}. Two major breakthroughs behind its popularity were the development of empirically based energy models---of which the most widespread is the Turner nearest neighbour energy model \cite{mathews:1999}---, and two fast dynamic programming algorithms to determine the minimum free energy (MFE) secondary structure \cite{zuker:1981} and to compute the partition function \cite{mccaskill:1990} of a sequence. In general, a sequence can fold into a number of secondary structures and the energy models and dynamic programming algorithms have made it possible to select low-energy structures \cite{wuchty:1999}, quantify their free energies \cite{lorenz:2011} and use this to define a GP map in several ways: one GP map definition considers a single structure per sequence, usually the minimum-free-energy structure \cite{schuster:1994}. This will lead to a many-to-one GP map, where each sequence maps to a single structure, but each structure can be generated by a number of different sequences. An alternative definition allows several low-free-energy structures per sequence, which leads to a more complex many-to-many relationship. Together, these different studies defined a range of formal measures to quantify some of the key features of GP relationships, such as plasticity, evolvability, robustness and modularity \cite{ancel:2000}. The results obtained with RNA through the years have served as inspiration and guide to our intuition when faced with other GP maps. 

\subsubsection{Phenotypic bias in RNA}
\label{sec:RNABias}
We will start by reviewing results from the commonly studied many-to-one GP map, where the focus is solely on the predicted minimum-free-energy structure of each sequence. The largest exhaustive enumeration performed for RNA sequences, of length $L=20$, yielded 10 orders of magnitude difference in the number of genotypes mapping from the most rare to the most frequent secondary structure phenotypes \cite{schaper:2014}. Approximate calculations of NSSs for longer sequences \cite{dingle:2015,garcia-martin:2018} show that this variance grows rapidly with increasing length. For example, for $L=100$ this difference is expected to be over 50 orders of magnitude: these maps are extremely biased. In an important study \cite{jorg:2008} the NSSs for longer length RNA were calculated using a sampling technique. When comparing to structures in the fRNAdb database for functional non-coding RNA (ncRNA) \cite{kin:2007}, they found, for systems of lengths $L=30$ to $L=50$, that the natural RNA secondary structures were typically among those with larger NSS. These results suggested that the strong bias in the GP map was reflected in the secondary structures found in nature. 

Another interesting set of studies compared structural features (e.g.\ distributions of stack and loop sizes) of natural secondary structures and those obtained when randomly sampling over sequences.  They found that many are quite similar \cite{fontana:1993}, and that natural and random RNA share strong similarities in the sequence nucleotide composition of secondary structure motifs such as stems, loops, and bulges \cite{smit:2006}. Why should random sampling over sequences generate distributions that are so similar to natural RNA, where natural selection would normally be thought to play an important role?

The study of much larger datasets of natural RNA from the fRNAdb database---and for lengths ranging from $L=20$ to $L=126$---demonstrated that the distributions of various structural features, and also properties such as the genotypic robustness, are very close to those obtained by random sampling over genotypes \cite{dingle:2015}. Furthermore, the distribution of NSS for natural RNA was found to closely follow the NSS distribution that arises upon random sampling of phenotypes. 
If one were to simply randomly sample over phenotypes, 
very significant differences with random genotype sampling (and natural RNA) would be found. By working out these counterfactuals it was therefore possible to demonstrate that the way in which variation arises through a GP map is dramatically different from the naive expectation that all potential variation is equally likely.
  
The close agreement of the distributions found in nature and those found by random sampling of genotypes via the GP map is very surprising given that natural selection is expected to be an important factor in the process that allows a particular functional RNA to fix in a population. The fact that its effect is not really visible for the properties above, at least when compared to a null model of random sampling genotypes, would appear to be strong evidence for the importance of anisotropic variation in determining evolutionary outcomes. However, before this conclusion can be drawn, it is important to remember that evolution does not proceed by random sampling of genotypes. Instead, it typically starts with a particular genotype and phenotype, and alters it via mutations that in turn generate new phenotypes that are either fixed or disappear over the generations in evolving populations. Given the hyper-astronomically large size of these spaces, it is not clear that such a local search should be at all similar to the results of random sampling of genotypes, which is a global property that does not depend on the starting point in genotype space.

Still, a counterexample of natural RNA where selection seems to have played a visible effect is that of viroids. Viroids are small, non-coding, circular RNA molecules that infect plants \cite{diener:1971}. Viroids have compact secondary structures that constrain their evolution \cite{elena:2009} and whose preservation seems essential to avoid degradation and inactivation \cite{diserio:2017}, and to minimise the effect of deleterious mutations \cite{sanjuan:2006:MBEI,sanjuan:2006:MBEII}. Viroids bear a number of paired nucleotides well above random expectations \cite{cuesta:2017}, such that the estimated NSSs of typical viroids are significantly below those of random sequences. For example, a typical structure for a circular RNA of length 399 has an average of 230 paired nucleotides and about $10^{91}$ compatible sequences. However, the largest known viroid is {\it Chrysanthemum chlorotic mottle viroid}, which matches that length, but has 280 paired nucleotides and an NSS of about $10^{72}$ genotypes \cite{catalan:2019a}.

\subsubsection{Promiscuity in RNA}

Beyond the many-to-one GP map, 
many-to-many GP maps that take into account the MFE structure and suboptimal structures in the Boltzmann ensemble have been studied \cite{ancel:2000, wagner:2014, rezazadegan:2018}. Suboptimal structures can be included according to several criteria: either all structures which fall within a fixed free energy range from the MFE structure \cite{ancel:2000, wagner:2014} are considered or only structures which have the same free energy as the MFE structure up to the energy resolution of the computational model \cite{rezazadegan:2018}. First, a link was found between the suboptimal phenotypes of a sequence in the many-to-many GP map and the phenotypes in the mutational neighbourhood of the same sequence in the corresponding many-to-one GP map \cite{ancel:2000}. Secondly, genotypes with low promiscuity were shown to have MFE structures with higher modularity \cite{ancel:2000}. Finally, it was found that evolving populations encounter a higher number of phenotypes if suboptimal phenotypes are included \cite{wagner:2014}. Altogether, these observations point to the important adaptive role of molecular promiscuity by supplying alternative phenotypes in the absence of mutations, and so redefining the fitness landscape \cite{aguirre:2018}.

\subsubsection{Hints from RNA inverse folding algorithms}

The characterisation of functional phenotypes by designing sequences that fold into a given RNA secondary structure has been much less explored than the direct fold of given sequences. Finding sequences that yield a particular secondary structure is known as the RNA {\it inverse folding} problem. This is an NP-complete problem even for the MFE structure \cite{schnall-levin:2008}, hence a very demanding computational task. As a consequence, most approaches are based on local search algorithms \cite{churkin:2017}. Actually, RNA inverse folding algorithms are mostly intended for synthetic design, though they have occasionally been used to investigate GP relationships \cite{wagner:2007,borenstein:2006}. However, their use is controversial due to the intrinsic bias of the underlying local search algorithms \cite{szollosi:2009}, which are not complete by definition and therefore produce biased samples under multiple runs. This caveat notwithstanding, there are some inverse folding methodologies that appear more suitable for this purpose. 

The first method is a {\em soft inverse folding} approach which implements a dynamic programming algorithm to compute the RNA {\em dual partition function} \cite{garcia-martin:2016b}. This partition function is defined as the sum of Boltzmann factors $\sum_{\sigma}\exp(-E(\sigma,\Sigma)/T)$, where $E(\sigma,\Sigma)$ is the energy of the RNA nucleotide sequence $\sigma$ compatible with a target structure $\Sigma$, and $T$ the absolute temperature (in units of energy). An energy weighted sampling from the low energy ensemble of sequences that are compatible with the given secondary structure is performed to calculate this partition function. While this approach is not particularly practical for synthetic design, it provides insights into molecular evolution.  

This theoretical abstraction and the measures derived from it, such as the {\em expected dual energy}, can provide useful information about general properties of the phenotypes without exploring the whole genotype space. Computational analyses based on the nearest neighbour energy model over all the RNA sequences in the Rfam database \cite{kalvari:2018} indicate that natural RNAs fold into secondary structures with energy higher than expected for sequences with the same length and GC content. Possible explanations for this observation are either that functional RNAs are not under evolutionary pressure to be highly thermodynamically stable or that sequence requirements prevent reaching minimum folding energies. On the other hand, experimental studies confirm that even random sequences frequently acquire compact folds similar to those of natural RNAs. Empirical observations further indicate that natural selection could be a determinant factor to achieve unique, stable tertiary folds---i.e. without major competing phenotypes---under natural conditions \cite{schultes:2005}. Besides, the controlled bias in this sampling methodology provides a delimited context to evaluate the properties that characterise a functional RNA with respect to sequences with similar structure. Simulations using this approach indicate that bacterial ncRNAs are more plastic and less robust than other sequences with similar structure \cite{garcia-martin:2016b}.

Although the samples returned by this algorithm are representative of the low energy ensemble of sequences of the given structure, the MFE structure of individual sequences is not necessarily the target structure. However, the proportion of alternative MFE structures of the sampled sequences is the distribution of {\em competing phenotypes} in the low energy ensemble of the target structure, which can in turn be interpreted as an estimate of the structures that are likely to coexist with that phenotype in a many-to-many GP map. 

Similar algorithms for computing and sampling from the {\em RNA dual partition function} with additional constraints have been developed and used to determine the neutral path between sequences in the same phenotype \cite{barrett:2018}.

The second methodology is complete inverse folding based on constraint programming \cite{garcia-martin:2013}. The constraint programming paradigm avoids exploring the whole sequence space when structural, sequence or environmental restrictions are included. These restrictions comprise, among many others, GC content, sequence motifs, multiple local and global structures and folding temperatures. Rather than slowing down the search, each constraint increases the speed of this algorithm. This algorithm can potentially retrieve all sequences that meet the requirements or conclude that no solution exists. In practice, the running time depends on the sequence space defined by the given constraints. These features make it appropriate for the study of genotype-phenotype-function relationships of moderately small functional RNAs with known moieties, or of regulatory RNA elements like riboswitches and thermoswitches.

Some examples of the performance of complete inverse folding based on constraint programming are the computationally-based suggestion that the conserved GUH (no G) motif in the hammerhead ribozyme type III cleavage site of {\it Peach latent mosaic viroid} is due to structural, rather than functional, requirements \cite{dotu:2014}, or that natural thermoswitches do not seem to be optimised to maximise the probability difference between the active and inactive structures at the corresponding folding temperatures \cite{garcia-martin:2016a}.

\subsection{Artificial life}

Evolutionary processes have not only been studied in biology, but also in man-made systems. Some models were designed to simulate biological evolution computationally and mimic biological properties. A widely used example is the digital model of a biological organism called Avida \cite{ofria:2004}. Avida organisms are pieces of code which can self-replicate and evolve towards optimal usage of computational resources. Richard Dawkins introduced a different form of artificial life to study evolution: biomorphs \cite{dawkins:2003} are two-dimensional stick figures produced recursively from a genotype, which consists of nine integer numbers. These biomorphs resemble abstract animal or plant shapes. Lindenmayer systems are another famous recursive model which can produce plant-like figures \cite{lindenmayer:1968a,lindenmayer:1968b}. These model systems are abstractions of biological organisms, but they all imitate properties of biological systems: the recursive branching rules in Lindenmayer's systems and later in Dawkins' biomorphs were inspired by plant development, whereas Avida digital organisms have a metabolism and compete, just like bacteria \cite{ofria:2004,lindenmayer:1968b,dawkins:2003}. However, evolutionary principles have been applied even more generally: the study of programmable electronic hardware has been addressed using the GP framework \cite{raman:2011}. Circuit configurations were treated as genotypes and the function which a circuit computes as the corresponding phenotype.

Here we will focus on results for four artificial life models: Avida organisms \cite{fortuna:2017}, biomorphs \cite{dawkins:2003,martin:2020}, the 2PD0L model \cite{lehre:2005,lehre:2007}, which is based on Lindenmayer's systems, and FPGAs \cite{raman:2011}, a type of programmable electronic circuits. These studies have focused on different properties, which makes a direct and quantitative comparison difficult. However, similarities between these artificial life GP maps and molecular sequence-to-structure GP maps exist \cite{fortuna:2017,lehre:2005,lehre:2007,raman:2011}: first, in three of these four systems the number of genotypes mapping to a given phenotype was estimated and found to vary significantly between phenotypes \cite{fortuna:2017,raman:2011,martin:2020}. For the fourth model a related quantity, the neutral set diameter, was also found to differ between phenotypes \cite{lehre:2007}. Such a heterogeneity, or phenotypic bias, in the distribution of genotypes over phenotypes has long been observed in molecular structure GP maps \cite{schuster:1994,li:1996}. Second, a high degree of genotypic robustness was observed, which enables the formation of  neutral networks \cite{fortuna:2017,raman:2011,lehre:2007,martin:2020}. This property was also first found in molecular structure GP maps \cite{lipman:1991} and is referred to as genetic correlations \cite{greenbury:2016}. A third shared property follows from the vastly different NSS: the probability of transitioning from a larger to a chosen smaller neutral set by point mutations is much smaller than that in the reverse direction. This asymmetry is known from molecular structure GP maps \cite{fontana:1998b} and has been confirmed for two of the artificial life GP maps: Avida \cite{fortuna:2017} and the 2PD0L model \cite{lehre:2005}.

In addition to these shared properties, there are points in which the various artificial life systems differ. In Avida, a high fraction of genotypes is considered inviable because the organisms are unable to reproduce \cite{fortuna:2017}, whereas in the biomorphs system all genotypes produce well-defined drawings and all stick figures are viable until an external decision is made about the fitness of specific shapes. In molecular GP maps the fraction of viable genotypes also depends on the system: in studies of model proteins, a large fraction of genotypes does not fold into a unique structure and is considered unstable, whereas for RNA secondary structure a minimum free energy structure is found for most sequences \cite{ferrada:2012}. Further comparisons could be made once quantities defined for GP maps, such as phenotypic robustness and evolvability, NSS, and mean-field mutation probabilities, are evaluated consistently for all of these and further artificial life models. Commonalities between artificial life and molecular structure GP maps dominate the picture at present, but future research may also identify differences between these two groups of models.

\section{The universal topology of genotype spaces}
\label{sec:UnivTopology}

Some of the results highlighted in the former section hint at the possibility that any sensible GP map (and, by extension, artificial life system) is characterised by a generic set of structural properties that appear repeatedly, with small quantitative variations, regardless the specifics of each map. Extensive research performed in recent years has confirmed this possibility to an unexpected degree. 

Some of the commonalities documented are navigability, as reflected in the ubiquitous existence of large neutral networks for common phenotypes that span the whole space of genotypes, a negative correlation between genotypic evolvability and genotypic robustness, a positive correlation between phenotypic evolvability and phenotypic robustness, a linear growth of phenotypic robustness with the logarithm of the NSS, or a near lognormal distribution of the latter. There are recent and comprehensive reviews of the properties measured and shared by different GP maps \cite{reidys:1997,stadler:2006,wagner:2011,ahnert:2017,aguirre:2018,nichol:2019}. In the following sections, we discuss new views on the plausible roots of this seemingly universal class of GP maps. 

\subsection{Possible roots of universality in GP maps}
\label{sec:PossibleRoots}

The question obviously arises: Why are structural properties of GP maps unaltered by the details of the mapping? Part of the answer must lie in the topology of the very high dimensional spaces governing the relationship between genotypes and phenotypes. Our intuitions often fail us here because these spaces are highly interconnected. Although their volumes grow exponentially with sequence length, distances are linear. For example, if one made every RNA of length $L=79$, the molecules would weigh more than the Earth \cite{louis:2016}. Yet none of those strands is more than 79 point mutations away from any other.  

One way this interconnection manifests itself is through the property of shape-space covering, a term first introduced for GP maps in the RNA context \cite{schuster:1994}, and borrowed from its original use in immunology \cite{perelson:1979}. It captures the fact that many phenotypes are only a handful of mutations away from one another. While this property has been best studied in the secondary structure RNA GP map, it has also been shown to be present in the HP model \cite{bornberg-bauer:1999,ferrada:2012}, toyLIFE \cite{catalan:2018}, the polyominoes \cite{greenbury:2014}, and a model of gene expression \cite{khatri:2009} (where it is described as ergodicity of phenotypic exploration). Shape-space covering suggests that no matter where you start, many other phenotypes are in principle close by in terms of Hamming distance. In the cases above, this holds even if the search begins in an arbitrary genotype. In GP maps where function is sparse in genotype space \cite{ciliberti:2007,matias-rodrigues:2009,barve:2013}, phenotypes are still close to each other, but links are established through a limited number of genotypes that might take a long time to find through random walks on the neutral network.

\subsection{Constrained and unconstrained sequence positions. Formalising neutrality and evolvability}
\label{sec:ConstrainedPositions}

The intuitions above have received quantitative support from  analytically tractable, streamlined GP maps which aim to capture the essentials of generic GP map features. These models, the results attained and the clues they provide are summarised in this section, which might appear slightly technical to the non-familiar reader but clarifies possible constructive principles of evolutionarily apt GP maps. 

Highly simplified, abstract GP maps can reproduce many of the generic properties discussed \cite{greenbury:2015,weiss:2018}. These simplified maps hint at two major possible causes underlying structural universality: (i) the partition of sequence regions into constrained and unconstrained parts and (ii) non-local interdependence of sequence positions with regard to their constraints (as sketched in Fig.~\ref{fig:GPmodels} (d,e)). Let us illustrate how to derive general results with a simple example. Consider a sequence of length $L$ whose first $\ell$ positions are fully constrained (changing any of those positions amounts to changing the phenotype) and the remaining $L-\ell$ positions are neutral (changes do not affect phenotype). If every position in the sequence admits $k$ possible values ($k=2$ for a binary alphabet, formed for example by symbols $\{0, 1\}$ and $k=4$ in quaternary alphabets, as $\{$A, C, G, U$\}$), then there are $k^{\ell}$ different phenotypes for every value of $\ell$, each of size $k^{L-\ell}$. Using a rank-ordering of the sizes of phenotypes and some simple algebra, it is easy to conclude that the probability $p(S)$ that a phenotype has size $S$ is $p(S) \propto S^{-\alpha}$, with $\alpha=2$ \cite{manrubia:2017}. If the restriction of a position being fully constrained or neutral is relaxed, different values of the exponent $\alpha$ can be obtained. The exponent also changes if a stop codon (equivalent to considering an $\ell$-dependent amount of lethal mutations) is introduced \cite{greenbury:2015}.
 
\begin{figure}[ht]
\begin{center}
\vspace{-0.3cm}
\includegraphics[width=6cm,clip]{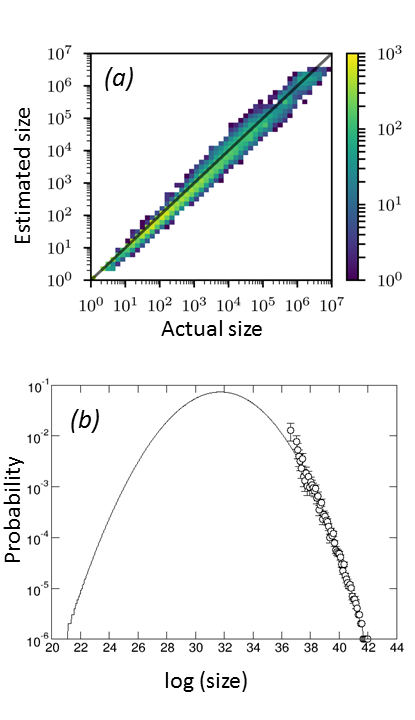}
\vspace{-1cm}
\end{center}
\caption{Predicted and measured properties of RNA phenotypes. (a) Log-log-log histogram of the estimated abundance vs. actual abundance of four-letter RNA of length $L=16$ phenotypes using versatility as defined in the main text \cite{garcia-martin:2018}; (b) NSSs estimated using \cite{jorg:2008} natural RNAs of length $L= 100$ obtained from the ncRNA database \cite{dingle:2015}. Random evolutionary search is highly skewed towards the largest phenotypes, as evidenced by the predicted shape of the full, lognormal distribution (solid curve): phenotypes of small and typical sizes are not found in nature.}
\label{fig:Versatility}
\end{figure} 
 
Interestingly, the shape of the distribution $p(S)$ changes to a lognormal function if, in the examples above, constrained sites can be arbitrarily distributed along the sequence. In general, the positions of a sequence are neither constrained nor neutral, but versatile in varying degrees. Let us define the {\it versatility} $v_i$ of position $i$ in a sequence as the average number of alphabet letters at that site that do not modify the phenotype. This extends the ideas above and provides a simple estimation of neutral set size $S$, as $S = v_1 v_2 v_3 \dots v_L$. This estimated value has been shown to be a very good approximation to the NSS in several GP models such as RNA, HP and toyLIFE \cite{garcia-martin:2018} (Fig. \ref{fig:Versatility}a). In all those cases and several others, the distribution $p(S)$ is compatible with a lognormal, which can be analytically derived under very generic assumptions in the case of RNA \cite{manrubia:2017} (Fig. \ref{fig:Versatility}b). Moreover, the results suggest that this approximation can be extrapolated to larger sizes. Additional properties, such as genotypic and phenotypic robustness, can be analytically obtained in such effective models \cite{greenbury:2015}, which constitute a sound first step towards deriving a formal theory of genotype spaces and their universal properties.
 
Generic biological sequences display the characteristics above in almost every biological context: exons and introns correspond to constrained and unconstrained regions, as do genes and noncoding intergenic sequences. Start and stop codons as well as interactions between transcription factors and their targets are examples of the interdependence of one sequence region on the constraint of another. As a result it is likely that the same GP map properties we observe in abstract model systems also hold for much more complex and biologically realistic phenotypes. The challenge in these more complex GP maps, however, is the vast size of the genotype space. A protein of 300 residues has a sequence space of size 20$^{300}$. Approaches that can estimate the structural properties of a GP map from relatively small samples are therefore essential. Knowledge of these properties is not just interesting for the study of GP maps, but also has potentially useful applications \cite{dingle:2015}. Being able to measure properties such as the phenotypic robustness, evolvability, and neutral network size of phenotypes in more complex GP maps would therefore provide a powerful methodological tool for the prediction of evolutionary pathways. The division of sequences into constrained and unconstrained regions is also likely to make prediction of structural GP map properties from local samples easier. This is because a division of sequences into constrained and unconstrained regions implies that many sequence positions are largely independent of each other with regard to their phenotypic effect. While important interdependencies remain, which particularly affect evolvability, the fact that interdependent sequence positions are likely to constitute a relatively small fraction of the total sequence means that a sampling approach is feasible for the purpose of estimating neutral network sizes and phenotypic robustness.

\section{Evolutionary dynamics on genotype spaces}
\label{sec:dynamics}

In the previous sections we have discussed the static properties of genotype spaces, their plausible universality and some basic principles that may underlie their topology. Such findings are relevant by themselves, but a further aim is to uncover the consequences of genotype space architecture in evolutionary dynamics. Evolution can be pictured as the navigation on the space of all possible genotypes \cite{maynard-smith:1970}, and GP maps describe the way different phenotypes are organised in such a space \cite{alberch:1991}. This organisation and the intrinsic structure of GP maps affects, among others, the ability to find genotypes and phenotypes in evolutionary searches \cite{schaper:2014,cowperthwaite:2008}, as well as the rate of adaptation \cite{draghi:2010,manrubia:2015}.

Early studies of dynamics on neutral networks quantified the trend of populations to maximise genotypic robustness by demonstrating that mutation-selection equilibrium is solely determined by the network topology \cite{nimwegen:1999}. Still, the time to reach equilibrium is an inverse function of the mutation rate \cite{aguirre:2009}. Neutral networks in GP maps, as well as in a few instances where this property could be quantified, are assortative \cite{aguirre:2011}: the neutrality of genotypes one mutation away from each other is positively correlated. As a result, the dynamics is naturally canalised towards maximally connected regions \cite{ancel:2000}, resulting in an acceleration in the rate of accumulation of neutral mutations with time \cite{manrubia:2015}.

In more recent analyses, attention has turned towards the effect of phenotypic bias in adaptation, as we have already discussed by means of enlightening studies with RNA. The question has been also investigated using a modified version of toyLIFE to model pattern-formation in regulatory networks \cite{catalan:2017tesis,catalan:2020} aimed at finding out how evolution chooses between two {\em a priori} equally fit phenotypes. It turns out that evolutionary dynamics at the phenotypic level cannot be well described by a Markovian process between phenotypes \cite{manrubia:2015}, because of the nontrivial topology of each phenotype's neutral network \cite{aguirre:2018}. As a matter of fact, the escape time from one phenotype does not follow an exponential distribution, as most evolutionary models assume. This is one instance of the so-called phenotypic entrapment \cite{manrubia:2015}, in which the trend of populations to become trapped in increasingly robust regions of a phenotype neutral network results in a long-tailed distribution of escape times: either the population escapes very fast, or takes a very long time to do it. 

Accounts of evolution on neutral networks driven by point mutations and the corresponding mathematical formalism can be found elsewhere \cite{wilke:2001BMB,reidys:2001,aguirre:2018}, though some essentials will be also described here. In this section we will mainly discuss the effects of a largely disregarded but essential evolutionary mechanism (recombination) and how mutational bias affects isotropic searches. 

We continue with evolutionary dynamics on genotype and phenotype networks defined by point mutations, where if we make an ergodic assumption that all typical phenotypes are locally accessible we are led in a natural manner to the formulation of the statistical mechanics of phenotypic evolution. We close by discussing a number of applications where these ergodic assumptions are most appropriate.

The approaches in this section differ in the formalism used (complex networks at large, mean-field effective models and statistical mechanics) but all converge in the main emerging lesson: the size of a phenotype plays a role in evolution comparable to that of fitness. Quantification of their relative weight through formal approaches might eventually settle the false dichotomy between neutralism and adaptationism. 

\subsection{Robustness and recombination}
\label{sec:Recombination}

Genotypic robustness is a property of the GP map that quantifies to what extent functional genotypes can be maintained in the presence of random mutations \cite{visser:2003,lenski:2006,masel:2010,wagner:2005}. Specifically, consider a genotype encoded by a sequence $\sigma$ of length $L$ that admits a total of $(k-1)L$ single point mutations (recall that $k$ is the size of the alphabet). Genotypes are classified to be either viable (functional) or lethal (non-functional). Then the genotypic robustness $r_\sigma$ of a viable genotype is defined as the fraction of mutations that maintain viability \cite{chen:2009}, $r_\sigma = n_\sigma/{(k-1)L}$, where $n_\sigma$ is the number of viable mutational neighbours. The population-averaged robustness is correspondingly defined as
\begin{equation}
\label{Krug_robustness}
r = \sum_{\sigma \in V} r_\sigma \nu_\sigma^\ast,
\end{equation}
where $V$ denotes the set of viable genotypes and $\nu_\sigma^\ast$ is the stationary frequency of genotype $\sigma$. Two limiting cases are of particular interest. If the product of population size $N$ and mutation rate $U$ per individual and generation is small, $N U \ll 1$, the population is monomorphic and performs a random walk on the network of viable states. The stationary frequency distribution is then uniform and (\ref{Krug_robustness}) reduces to
\begin{equation}
\label{Krug_uniform}
r_0 = \vert V \vert^{-1} \sum_{\sigma \in V} r_\sigma,
\end{equation}
where $\vert V \vert$ is the number of viable genotypes. On the other hand, when $NU \gg 1$, the stationary frequency distribution is determined by mutation-selection balance and can be shown to be given by the leading eigenvector of the adjacency matrix of the network of viable genotypes \cite{bornberg-bauer:1999,nimwegen:1999}, see also Section~\ref{sec:PhenotypicTransitions}. The population robustness $r$ is related to the corresponding eigenvalue and exceeds the uniform robustness $r_0$ whenever the network is inhomogeneous. This implies that selection in large populations increases robustness by focusing the population in highly connected regions of the network.

Numerical studies of recombining populations on various types of genotype networks have indicated that recombination enhances the focusing effect of selection and thus substantially increases genotypic robustness \cite{azevedo:2006,hu:2014,huynen:1994,singhal:2019,szollosi:2008,xia:2002}. Recently, a systematic and largely analytic investigation of the relationship between recombination and genotypic robustness within the framework of deterministic mutation-selection-recombination models has been presented \cite{klug:2019}.

As a simple but informative example, consider the space of binary sequences $\{0,1\}^L$ endowed with a `mesa' landscape where genotypes carrying up to $\eta$ 1's are viable and all others are lethal \cite{wolff:2009}. The genotypes on the brink of the mesa carry exactly $\eta$ mutations and have robustness $r_\sigma = \eta/L$, whereas all others have robustness $r_\sigma = 1$. Combinatorial considerations show that the uniform robustness $r_0 \approx 2\eta/L$ for large $L$ and $\eta < L/2$, reflecting the fact that a large fraction of genotypes are located at the brink for purely entropic reasons. The maximal robustness that can be achieved through selection alone is \cite{wolff:2009}
\begin{equation}
\label{Krug_kL}
r \approx 2 \sqrt{\frac{\eta}{L}\left( 1 - \frac{\eta}{L} \right)} \;\; \textrm{for} \;\; \eta < L/2,
\end{equation}
which exceeds $r_0$ but is small compared to unity when $\eta \ll L$. Thus selection only partly counteracts the entropic outward pressure and as a consequence a large part of the population is still located near the brink under mutation-selection balance. By contrast, in the presence of recombination $r \to 1$ for small mutation rates, because the contracting property of recombination efficiently transfers the population to the interior of the mesa where all genotypes are surrounded by viable mutants.  

\begin{figure*}[t]
\begin{center}
\includegraphics[width=\textwidth]{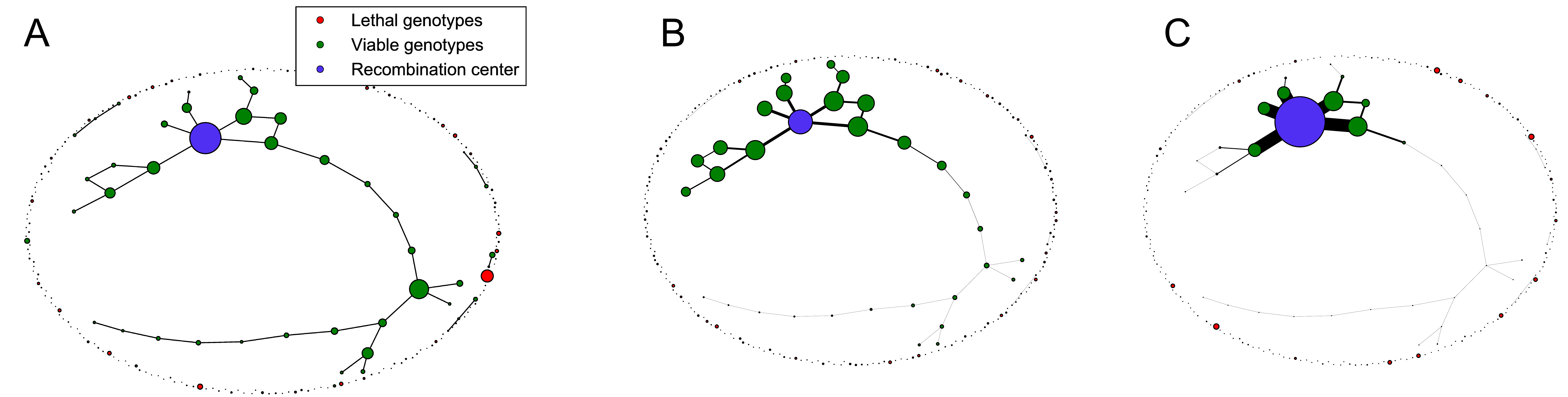}
\end{center}
\caption{\label{JK-Fig2} Genotype network generated by assigning viable genotypes at random with probability $p=0.2$ to binary sequences of length $L=8$. The largest connected component of viable genotypes is shown in the centre of each panel, and smaller components and isolated nodes are arranged in a ring surrounding the central component. (a) Network structure visualised by the recombination weight $\omega_\sigma$. Node areas are proportional to $\omega_\sigma^6$ and the recombination centre is marked in purple. (b) Stationary frequency distribution of a  non-recombining population. Node areas are proportional to the stationary frequency $\nu_\sigma^\ast$ of the respective genotype, and the edge width between neighbouring genotypes $\sigma$, $\tau$ is proportional to $\max[\nu_\sigma^\ast, \nu_\tau^\ast]$. (c) Same as (b) for a recombining population. Note that the population is much more strongly concentrated on the recombination centre than in panel (b). In panels (b) and (c) the mutation rate per site is $\mu = U/L = 0.001$. Courtesy of Alexander Klug.}
\end{figure*}

Simulations on different types of random genotype networks show that the massive enhancement of robustness found for the mesa landscape is generic, and typically a recombination rate on the order of the mutation rate suffices to achieve this effect. It is not obvious that the focusing of the population towards the centre of its genotypic range by recombination should generally increase robustness in this case, because viable and lethal genotypes are randomly interspersed in the network. To rationalise the observed increase in robustness it is useful to quantify the likelihood of a genotype $\sigma$ to be created by recombination through its \textit{recombination weight} $\omega_\sigma$ defined by
\begin{equation}
\label{Krug_recombination_weight}
\omega_\sigma = \frac{1}{\vert G \vert} \sum_{\kappa \in V, \tau \in V} R_{\sigma \vert \kappa \tau}.
\end{equation}
Here $R_{\sigma \vert \kappa \tau}$ denotes the probability that $\sigma$ is generated by crossover from $\kappa$ and $\tau$ and $\vert G \vert$ is the total number of genotypes. The normalisation ensures that $\omega_\sigma \in [0,1]$, and the recombination weights sum to $\sum_\sigma \omega_\sigma = \vert V \vert^2/\vert G \vert$. The genotype that maximises $\omega_\sigma$ is called the \textit{recombination centre} of the network and provides a good predictor for the point of concentration of the recombining population in the limit $U \to 0$ (see Fig.~\ref{JK-Fig2} for an example). Moreover, for two classes of random, percolation-type genotype networks and one empirical fitness landscape, the recombination weight $\omega_\sigma$ was found to be positively correlated with the genotypic robustness $r_\sigma$. 

If this correlation were a generic feature of GP maps, it would constitute a mechanistic explanation for how recombination acts to enhance genotypic robustness. Future work should therefore elucidate the conditions on the topology of the genotype network required for such a correlation to be present. It is not difficult to construct counterexamples where the recombination centre has low robustness, e.g., by placing a hole of lethal genotypes at the centre of a mesa landscape. Only the investigation of specific, biophysically motivated GP maps such as RNA secondary structures or lattice proteins will clarify whether or not such instances are statistically relevant. More broadly, it appears that a common perspective on recombination, robustness and evolvability \cite{masel:2010,lenski:2006,wagner:2005} may help to develop and test novel hypotheses about the evolutionary origins of these important biological phenomena.    

\subsection{Mutation bias}

Some regions of genotype space exhibit biases in the mutations they contain. For instance, GC-rich regions have more G$\leftrightarrow$C transversion (purine-to-pyrimidine or pyrimidine-to-purine) mutations than transitions (pyrimidine-to-pyrimidine or purine-to-purine mutations). This may interact with biases in the generation of genetic variation, because some mutations occur more frequently than others. For instance, the rate of A$\leftrightarrow$G transitions exceeds the rate of T$\leftrightarrow$C transitions in transcribed human genes, whereas there is no significant difference in non-transcribed regions \cite{green:2003}. Furthermore, CpG dinucleotides---regions of DNA where C follows G---are considered ``hot spots'' for G$\rightarrow$A and C$\rightarrow$T transition mutations \cite{nachman:2000}. Other forms of mutation bias such as deletion bias and strand-specific bias have been reported in bacterial genomes \cite{paul:2013,mira:2002}.

Under certain population genetics conditions, mutation bias can be a orienting factor in adaptive evolution \cite{mccandlish:2014,stoltzfus:2017}, and several experimental evolution studies indicate that mutation bias can influence trajectories of adaptive protein evolution \cite{lozovsky:2009,rokyta:2005}. It is possible to get a better understanding of how such mutation biases affect the outcomes and mutational trajectories of adaptive evolution by studying their impact on the navigability of GP maps.

Instead of the classic depiction of a GP map in which all the possible mutations are equally likely to occur, one could consider regions of the genotype space being differentially prone to distinct kinds of mutations. Ultimately this would affect the probability of traversing different edges in the genotype network and, therefore, its navigability. In this context, a {\it mutation bias weight} could be formally defined and introduced into a more general formulation of genotype networks, by biasing the accessibility of different genotypes. Understanding the potential evolutionary implications of mutation biased GP maps could provide us with valuable information about the nature of the systems they represent. For example, if a bias towards certain kinds of mutations enhances the ability to find the adaptive peaks of a certain GP map, a testable prediction could be that adaptive genotypes are more likely to evolve in regions of the genome that are prone to that particular kind of mutation. 

Moreover, integrating mutation bias into the study of GP maps can change properties such as robustness and evolvability \cite{cano:2020,sella:2005}. Both robustness and evolvability are based on the structure of genotypic neighbourhoods, and this structure can change if mutation bias is considered. For instance, a genotype might seem highly robust when most of its neighbours in the genotype space map onto the same phenotype. However, if there is a sufficiently high mutation bias towards mutations that do not preserve that phenotype, robustness would be diminished. The same principle can apply to evolvability.

\subsection{Phenotypic transitions as competitions between networks}
\label{sec:PhenotypicTransitions}

\begin{figure*}
\begin{infobox}[frametitle= Box~\ref{box:networks}: Genotype spaces as networks of networks]
{Populations evolve in steadily changing environments where the impact of internal and external perturbations can rarely be considered in full. Often, nonlinear responses to small external changes hinder predictability, as weak perturbations might trigger critical transitions that strongly influence the fate of whole ecosystems \cite{may:1977,scheffer:2001}. Complex network theory and the tools associated to it offer a powerful framework to tackle this type of dynamical systems, since a multitude of natural systems can be modelled as nodes (agents) connected by links (interactions). 

While network science has largely focused on single networks, in the last decade the study of dynamical properties on networks of networks or, in a more general way, on multilayer networks \cite{kivela:2014}, has attracted wide attention \cite{gao:2011,quill:2012}. One important motivation has been the finding that robustness, synchronisation or cooperation lead to different behaviour when studied in isolated or in interconnected networks \cite{buldyrev:2010,aguirre:2014,gomezgardenes:2012,iranzo:2016}. However, the main reason for this change of perspective has been to realise that many natural systems, beyond displaying a network-like organisation, are also made of interacting and competing networks at very different scales, from the molecular level to supranational organisations \cite{buldu:2019}.

The extent to which network science can foster our knowledge and comprehension of the evolution and adaptation of heterogeneous populations in an ever changing biosphere is a relevant open question. In particular, the {\it theory of competing networks} can be used to analyse the evolutionary dynamics of populations in a space of genotypes that can be regarded as a network of networks \cite{yubero:2017}. From this viewpoint, population evolution is described as a competition for resources of a certain kind, where the competitors are whole networks instead of independent nodes \cite{aguirre:2013}.
}

\label{box:networks}
\end{infobox}
\end{figure*}

Formal studies of the way the structure and navigability of GP maps affects evolutionary dynamics can provide insights into the mechanisms underlying adaptive evolution, robustness and the emergence of phenotypic innovations. In the previous two sections, it has been shown that links between genotypes in a genotype network are weighted: microscopic mechanisms such as recombination and mutation bias modify the likelihood of transitions between pairs of genotypes. Constant link weights of a generic transition matrix {\bf M} correctly describe mutation bias, but cannot account for the effects of recombination, since in the latter case they depend on the abundances of each genotype $\nu_\sigma$ in a nonlinear way, and in general are a time-dependent quantity. The simultaneous consideration of point mutations and recombination in a network framework remains as a topic for future studies. 

In the following, we summarise a mutation-selection evolutionary process on a network of genotypes subject only to point mutations using tools from complex network theory. Consider a vector $\vec n(t)$ whose components are the population of individuals at each node at time $t$ (upon normalisation, each component $n_{\sigma}(t)$ is the frequency of the genotype $\nu_{\sigma}(t)$). Then,
\begin{equation}
\label{eq:M*n}
\vec n(t+1)=\textbf{M} \vec n(t)
\end{equation}
represents the dynamics of the population, where {\bf M} is a transition matrix with information on the fitness of each genotype, on the mutation and replication process, and on the weighted topology of the network. $\vec n(t)$ describes the distribution, at each time $t$, of the population of sequences on the space of genotypes. As already stated, mutation-selection equilibrium is independent of the initial state and given by the eigenvector $\vec u_1$ associated to the largest eigenvalue $\lambda_1$ of {\bf M}. Furthermore, $\lambda_1$ yields the growth rate of the population at equilibrium, and $\vec u_1$ is also a measure (known as eigenvector centrality) of the topological importance of a node in a network \cite{newman:2010}.

In the context of the theory of competing networks, any dynamics that takes place on networks interconnected through a limited number of links (networks of networks), can be often characterised as a competition where the contenders are whole networks, and where eigenvector centrality represents the resource that the agents compete for (see Box~\ref{box:networks}). The final outcome of such a struggle for centrality strongly depends on the internal structure of the competing networks and on the links connecting them \cite{aguirre:2013}.

On the other hand, it has been shown \cite{aguirre:2015} that even when environmental perturbations are weak, populations may suffer critical transitions in their genomic composition when the fraction of lethal mutations (i.e.~of zero-fitness genotypes) is sufficiently high---of the order of that observed in natural populations \cite{eyre-walker:2007}. A recent analysis of these results suggested that the space of genotypes can be regarded as a network of networks in ``competition'' to attract population \cite{yubero:2017}, and that knowledge of the topology of the space of genotypes entails a certain predictive capability of the future evolutionary dynamics of the population under study. In fitness landscapes with a large fraction of lethal genotypes (as it could be the case of the non-compact HP model, GP maps for gene regulatory networks, or models for metabolism), the space of genotypes is formed by many subnetworks connected through narrow adaptive pathways. This topology induces drastic transitions of population from one subnetwork to another, occasionally causing the extinction of the population. The key topological element underlying sudden genomic shifts is the high heterogeneity in the network describing and linking viable genotypes. This topology can arise under a significant fraction of lethal mutations (or non-viable genotypes), but the same phenomenon is observed in rugged fitness landscapes. 

\subsubsection{An empirical test of the theory: transition forecast}
\label{sec:empiricaltest}

It is highly likely that large molecular populations able to evolve fast, such as RNA viruses, can provide an empirical test of this predicted critical behaviour. The enormous advances of high-throughput sequencing allow for a very precise description of the populations at the molecular level, and in particular of the abundances of the coexisting genotypes. This information might be used to build the space of sequences associated to a population that evolves in a changing environment, and thus a proxy of the network of genotypes where the population evolves. Applying the theory of competing networks it is conceivable that the eigenvalues of the different subnetworks and the centrality of the connector nodes would provide valuable information on how environmental variability affects the sharpness of the transitions and on the chances that the population could survive. The combination of tools from complex networks theory and the last decades' research on state shifts in the biosphere \cite{barnosky:2012,brook:2013} might eventually lead to a prediction of the time left until the transition occurs. This prediction is important because, once a tipping point takes place, it becomes very difficult, if not impossible, to return to the previous state. At present, a wide variety of early warning signals for state shifts has already been characterised, but none of them yields precise information about the time left before the tipping point is reached \cite{scheffer:2009,scheffer:2012}. However, calculations of the minimal distance between the first and second eigenvalues associated to the transition matrix {\bf M} could be used to obtain a first estimation of the time to the transition \cite{aguirre:2015}. In an evolving population, the relative abundances of the different genotypes could be used as an approximation of the eigenvector $\vec{u}_1$; a measure of the growth rate of the population at equilibrium could yield the largest eigenvalue $\lambda_1$, and $\lambda_2$ might be estimated by quantifying how resilient the population is to external perturbations \cite{dai:2012}. A sufficiently precise measurement of these quantities would represent a very fruitful connection between actual evolving populations and a dynamical description of possible sudden evolutionary transitions. On a related note, regarding the space of genotypes as a network of networks entails a more coarse-grained, effective model where each genotype network can be considered as a single node, and where the dynamics can be simplified to account only for changes in the phenotype. Links in this higher-level description would have a weight proportional to the within- and between-phenotypes links. At odds with the description at the genotype level though, transitions between phenotypes are no longer symmetrical \cite{fontana:2002,cowperthwaite:2008}, nor is the dynamics describing these transitions Markovian any more \cite{huynen:1996,manrubia:2015}.

\subsection{A mean-field description of phenotype networks}

The qualitative properties of a high-dimensional evolutionary search are inherent to navigable GP maps and very likely responsible for some of the generic features described in Section~\ref{sec:UnivTopology}. Despite all caveats that the complex dynamics at the genotype level may raise due to its non-Markovian nature \cite{huynen:1996,manrubia:2015}, the high dimensionality of genotype spaces helps us understand why a simple mean field model \cite{schaper:2014}, which averages over much of the local structure of a neutral set, succeeds in capturing some of those generic, dynamical properties. The model works with $\phi_{\xi \chi}$, the probability that a point mutation for genotypes that map to phenotype $\xi$ generates a genotype for phenotype $\chi$, averaged over all genotypes that generate $\xi$. By measuring the $\phi_{\xi \chi}$, a weighted network between all the phenotypes can be defined, with $\phi_{\xi \chi}$ as the weights. This allows for a much simpler dynamics that ignores the individual genotypes, and so analytic results can be derived for many properties in dynamical regimes ranging from the monomorphic to the fully polymorphic limits. Interestingly, for RNA, as well as for a number of other GP maps \cite{greenbury:2016}, it was found to a good first approximation that if $\xi \neq \chi$ then
\begin{equation}
\phi_{\xi\chi} \approx f_\chi,
\end{equation}
where $f_\chi$ is the global frequency of phenotype $\chi$, i.e., the fraction of genotypes that map to $\chi$. Since the $f_\chi$ range over many orders of magnitude, so do the $\phi_{\xi\chi}$. In contrast to the case where $\xi\neq\chi$, the robustness of phenotype $\xi$ is $\phi_{\xi\xi} \propto \log(f_\xi)$, and so varies much less with NSS. This property of the robustness is critical for neutral exploration. The mean field model predicts that for many different starting phenotypes $\xi$, the probability that a different phenotype $\chi$ will appear as potential variation will scale as $f_\chi$.

For several GP maps, this simplified model does an excellent job at predicting the rates at which variation arises in full GP map simulations. Since NSS, or equivalently $f_\chi$, varies over many orders of magnitude, this argument predicts that, to first order, the rate at which variation arises will also vary over many orders of magnitude. Therefore, even though the set of physically possible variations may be very large, only a tiny fraction of the most frequent phenotypes will ever be presented to natural selection. This \textit{arrival of the frequent} effect \cite{schaper:2014} is therefore very strong. Fundamentally it is a non-steady state effect, since the longer an evolutionary run proceeds, the more the potential variation with lower $f_\chi$ becomes likely to appear. The arrival of the frequent differs from the {\em survival of the flattest}, \cite{wilke:2001Nat} which describes the situation where a fitness peak with lower fitness can nevertheless dominate over a higher fitness peak with a lower NSS. The latter effect can be analysed in a steady-state framework, whereas the former effect cannot. 

Let us return in this context to the question of why so many structural features, as well as the genotypic robustness of RNA secondary structures, are so accurately predicted by a null model that ignores selection entirely. The arguments above suggest that even in the more complex situation of RNA evolution in nature, variation will nevertheless to a good first approximation arise with a probability proportional to its NSS. Since this rate varies by so many orders of magnitude, this arrival of the frequent effect determines what natural selection can work with, and so tends to dominate over local fitness effects. Rare phenotypes will almost have no bearing on evolutionary dynamics: they will hardly be found by a population searching for an adaptive solution and, if they are found, they will be quickly lost due to mutations. This is akin to an entropic effect in statistical physics: dynamics tend to favour macrostates with a larger set of microstates.

In other words, natural selection can only act on variation that has been pre-sculpted by the GP map. For the case of RNA described in Section~\ref{sec:RNABias}, it appears that it mainly works by further refining parts of the sequence. This picture of the primacy of variation stands in sharp contrast to more traditional arguments about the importance of natural selection as an ultimate explanation of any evolutionary trends. It also raises many open questions. Are there other GP maps for which we can see such dramatic effects in nature? There are certainly conditions where this primacy of variation is incorrect.  But how, and when does this GP map based picture of pre-sculpted variation break down? The exceptional case of viroids, discussed in Section \ref{subsec:RNA} might be one such example, and provide clues to seek answers to the latter question. 

\subsection{Equilibrium properties and statistical mechanical analogies in the weak mutation regime}
\label{sec:SMevol}

The broad question of optimisation in evolution, such as the existence of a Lyapunov function, describing a general dynamics and approach to equilibrium was first addressed by Iwasa \cite{iwasa:1988} in his definition of ``free fitness'', in analogy to the free energy of statistical mechanics, and then later rediscovered for the particular case of the weak mutation regime \cite{sella:2005} and in the context of the evolution of quantitative traits \cite{barton:2009,barton:2009a}. The key insight, is that, at finite population size, not fitness itself but a combination of fitness and Shannon entropy (weighted by $1/N_e$, where $N_e$ is the effective population size) is optimised over the evolutionary degrees of freedom of interest. This perception is consistent with the mean-field description reviewed in the previous section, where it has been shown that phenotypic bias is at least as relevant as phenotype fitness in evolutionary dynamics. 

From a statistical mechanics viewpoint, and in the weak mutation regime ($N_eU\ll1$, $N_eU\ln(N_es)\ll1$, where $s$ is the gain in fitness brought about by a mutation), populations are approximately monomorphic and the degrees of freedom of interest are the different alleles, codons or genotypes, which are fixed, or not, in the population; evolution can be described by a Markov process, where populations effectively jump sequentially between adjacent genotypes by a substitutional process \cite{mccandlish:2014}, where in equilibrium (assuming uniform mutation and the fixation probability given by the diffusion approximation) the probability of occupation is given by the Boltzmann distribution
\begin{equation}\label{Eq:GenotypeBoltzmann}
p_\sigma = e^{2N_eF_\sigma}/Z,
\end{equation}
where $F_\sigma$ is the fitness of genotype, allele, or codon $\sigma$, and $Z$ is the partition function. It is clear that $N_e$ plays the role of inverse temperature, such that fitness dominates at large population sizes (low temperature) and genetic drift for small population sizes (high temperature). Many of the calculational tools of statistical mechanics and generating functions then carry over to evolutionary problems \cite{barton:2009a} under usual ergodic assumptions.

The statistical mechanical analogy finds particular use in understanding the evolution of phenotypes arising from GP maps. Here, selection acts on phenotypes, but mutation and variation arise at the level of genotypes. Keeping in mind the many-to-one nature of most GP maps and phenotypic bias, the Boltzmann distribution of genotypes can be recast in terms of a Boltzmann distribution of phenotypes \cite{khatri:2009,khatri:2015}; as each genotype mapping to a given phenotype must by definition have the same fitness, the probability of each phenotype is the Boltzmann factor ($e^{2N_eF(\xi)}$) weighted by the degeneracy of each phenotype $\Omega(\xi)=k^L f_\xi$, giving
\begin{equation}\label{Eq:PhenotypicBoltzmann}
p(\xi) = e^{2N_e\Phi(\xi)}/Z,
\end{equation}
where
\begin{equation}\label{Eq:FreeFitness}
\Phi(\xi) = F(\xi)+\frac{S(\xi)}{2N_e}
\end{equation}
is the free fitness of phenotypes, $S(\xi)=\ln(\Omega(\xi))$ being the Boltzmann or \textit{sequence entropy} of phenotypes. We see that for small populations phenotypes with larger sequence entropy are favoured by genetic drift in evolution.

\subsubsection{Statistical mechanics of the evolution of transcription factor-DNA binding}

The ideas above first formally found application in simple biophysical models of transcription factor DNA binding \cite{berg:2004,mustonen:2005,mustonen:2008}, where the degeneracy of binding affinities can be exactly quantified under simplifying assumptions. It is typically found that for a given transcription factor, the amino acids at the binding interface tend to have strong preference to bind a single nucleotide; it is then mismatched nucleotides that control the binding affinity, as these are strongly destabilising, since hydrogen bonding is disrupted at the interface, as well as the lost hydrogen bonds with water molecules. A simple model of transcription factor binding, assumes binding between protein and DNA can be reduced to either a quaternary or binary alphabet, where the binding energy $E$ is proportional to the number of mismatches, or Hamming distance, $h$, $E=\epsilon h$. The degeneracy function is then related to the binomial coefficient
\begin{equation}\label{Eq:BinomialDegeneracy}
\Omega(h) \propto \binom{L}{h}(k-1)^h.
\end{equation}

This simple combinatorial argument shows that there is a huge degeneracy, or phenotypic, bias towards poor binding in this genotype-phenotype map. This methodology has been used to infer the effective genome-wide fitness landscape for transcription factor DNA binding in {\it Escherichia coli} and yeast \cite{mustonen:2005,mustonen:2008,haldane:2014}, suggesting that on average binding is under stabilising selection, with monotonically decreasing fitness with decreasing binding affinity.

This simple model of transcription factor DNA binding suggests that smaller populations bear a significantly greater drift load under stabilising selection, than would be predicted if we assumed evolution based on phenotypes only \cite{khatri:2015,khatri:2015a}; while selection pushes populations to larger binding affinities, there is an opposing sequence entropic pressure for poorer binding. In equilibrium, these opposing tendencies are balanced, and it is the free fitness that is maximised, not fitness. This effect of sequence entropy on drift load is significantly greater than would be expected for a trait under stabilising selection, which ignores any degeneracy (see Box~\ref{box:quadraticlandscape}).

\begin{figure*}
\begin{infobox}[frametitle= Box~\ref{box:quadraticlandscape}: Quadratic free fitness landscape]
{We examine a quadratic free fitness landscape, which is a simple description for the statistical mechanics of transcription factor DNA binding. For moderately large $L$, we can make the standard Gaussian approximation of the binomial distribution in the degeneracy function \eqref{Eq:BinomialDegeneracy}, such that the sequence entropy function (up to a constant) is approximately quadratic:
\begin{equation}\label{Eq:QuadraticSequenceEntropy}
S(h) \approx \frac{k}{2\langle h \rangle}(h-\langle h\rangle)^2,
\end{equation}
where $\langle h\rangle = (k-1)L/k$ is the mean of the binomial distribution ($\langle h\rangle=L/2$ for binary, and $\langle h\rangle=3L/4$ for quaternary alphabets). Further, if we assume that the fitness landscape of binding affinities is quadratic of the form $F(h)=-\frac{1}{2}\kappa_Fh^2$, whose maximum is for the best binders ($h=0$), then from Eq.~\eqref{Eq:FreeFitness} the free fitness of binding energies/Hamming distance $h$ is then itself quadratic with new curvature $\kappa=\kappa_F+\frac{k}{2\langle h\rangle N_e}$ and with shifted maximum $h^*=\frac{k}{2N_e\kappa}$:
\begin{equation}\label{Eq:FreeFitnessTF}
  \Phi(h) = \frac{1}{2}\kappa(h-h^*)^2.
\end{equation}
This new maximum is shifted to poorer binding affinities and represents the balance between selection and sequence entropy:
\[
\frac{\ud\Phi}{\ud h} = \frac{\ud F}{\ud h} +\frac{1}{2N_e}\frac{\ud S}{\ud h} =0.
\]
It is instructive that the drift-load for this simple GP map varies as $D\sim N_e^{-1}$, a far stronger dependence on population size than if we considered evolution on only a phenotypic landscape $F(r)$, which would vary as $D\sim N_e^{-1/2}$; this significant difference arises as the sequence entropic pressure causes the peak of the phenotypic distribution to shift, whilst ignoring this would simply give rise to a broadening of the distribution.
}
\label{box:quadraticlandscape}
\end{infobox}
\end{figure*}

\subsubsection{Evolution of genotypic divergence and reproductive isolation}

\begin{figure*}[htbp]
\includegraphics[width=0.65\textwidth]{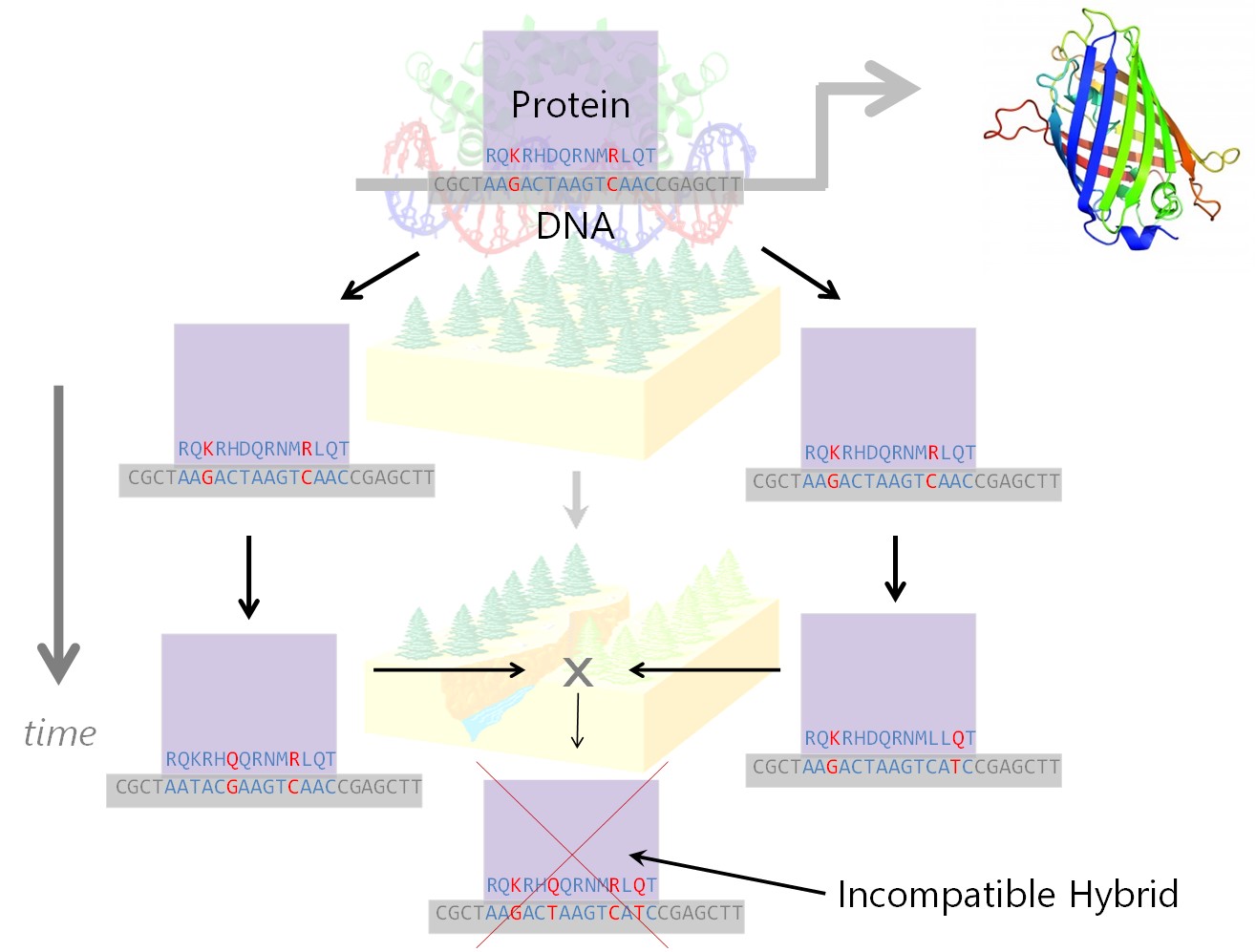}
\caption{Divergence for a simple GP map of transcription factor DNA binding. After a geographic split a once unique species evolves into two independent ones. Within simulations the fitness of various hybrid combinations of loci from each lineage can be calculated and number of inviable combinations (incompatibilities) recorded. Numerical results show that incompatibilities arise more quickly in smaller populations \cite{khatri:2015a}.}
\label{fig:GPmap}
\end{figure*}

One consequence of this significantly larger drift load is that in (allopatrically) diverging populations this gives rise to the prediction that reproductive isolation arises more quickly due to common ancestors already having more maladapted transcription factor-binding site pairs on average \cite{tulchinsky:2014,khatri:2015,khatri:2015a} (Fig.~\ref{fig:GPmap}); if the common ancestor has a binding affinity closer to being deleterious (but kept in check by stabilising selection) then in hybrids Dobzhansky-Muller incompatibilities \cite{dobzhansky:1936,muller:1942,bateson:2009}, which are incompatible combinations of transcription factors and DNA binding sites, arise more quickly after divergence. In particular, this mechanism is broadly consistent with trends seen in field-data \cite{fitzpatrick:2004,stelkens:2010,cooper:1997} and diversification rates in phylogenetic trees \cite{coyne:1998,barraclough:2001,nee:2001}, and so gives a robust explanation of how stabilising selection can give rise to this population size effect in speciation, without requiring passing through fitness valleys as do models based on the founder effect \cite{lande:1979,lande:1985,barton:1984,barton:1987}. This model also predicts that those transcription factor-DNA binding site pairs, which are under weaker selection across a genome, would for the same reason give rise to a greater contribution to reproductive isolation, as the balance between selection and sequence entropy would be shifted to give common ancestors with weaker binding on average \cite{khatri:2015a}.

\subsubsection{Marginal stability of compact proteins}

Equilibrium statistical mechanical ideas also have the potential to explain the observed marginal stability of compact proteins. Various databases show that proteins have stabilities (measured through free-energy differences) of order $\Delta G\sim-10$ kcal/mol, which is only a few hydrogen bonds \cite{zeldovich:2007}, when their potential maximum stability could be orders of magnitude greater. Although adaptive explanations have been suggested related to the necessity of protein flexibility \cite{zavodszky:1998}, a more straightforward explanation is offered in light of the free fitness of phenotypes; for a given chain length there are many more sequences that give poor protein stability and so this sequence entropy pressure balances the tendency of natural selection to choose proteins of higher stability. Simulations and theory of the evolution of protein folding \cite{taverna:2002,bloom:2007,zeldovich:2007,goldstein:2011,serohijos:2012} reproduce this marginal stability, with a particular property that the distribution of $\Delta\Delta G$, the change caused by mutations in the stability of a protein, is roughly independent of $\Delta G$, the stability of the protein. The marginal stability of proteins shows an interesting behaviour as function of population size $N_e$; as we might expect, as the population size is increased, selection dominates genetic drift, and simulations \cite{goldstein:2011,wylie:2011} show that the average stability 

\begin{equation}
\langle\Delta G\rangle\sim -k_BT\ln(N_e)    
\label{eq:scaling}
\end{equation}
in the weak mutation limit ($\mu N_e\ll 1$). This result can be obtained from simple scaling arguments \cite{wylie:2011}, and it has been theoretically shown that, under global selection against misfolding, a broader scaling relationship between protein folding stability, protein cellular abundance, and effective population size holds \cite{serohijos:2013}. Also, Eq. (\ref{eq:scaling}) can be shown to arise from similar reasoning to that in Box~\ref{box:quadraticlandscape}, as a result of a balance between selection and sequence entropy enhanced genetic drift at a given population size $N_e$ \cite{khatriUnpub:2019}.

\subsubsection{Free fitness, universality and developmental system drift}

The question naturally arises: how universal or general is this effect in the genetic divergence of populations under stabilising selection? In developmental biology it is commonly found that closely related species that have similar organismal level phenotypes, such as body plans, nonetheless have diverged in the regulatory networks that control this patterning \cite{matute:2010,verster:2014,wotton:2015}. This cryptic variation is known as ``developmental system drift'' \cite{true:2001,haag:2014} and is a potential source of hybrid incompatibilities and ultimately reproductive isolation as previously explored by using simple gene regulatory networks \cite{johnson:2000,johnson:2001}. However, in analogy to transcription factor DNA binding, if the GP map of developmental patterning has large degeneracy or phenotypic biases, we may also expect to find a rapid increase in the rate that hybrid incompatibilities arise as the population size decreases.

To explore this, a previously studied multi-level GP map for developmental spatial patterning \cite{khatri:2009} was used, in which gene regulation is manifested by multiple transcription factor DNA binding interactions, each described by a Hamming distance model as described above. Importantly, the GP map has the property that stabilising selection acts to maintain a body patterning phenotype, but the underlying genotypic degrees of freedom and molecular (binding energy) phenotypes can drift. In addition, the GP map has the essential property that allows an equilibrium analysis \cite{khatri:2009} that it is ergodic for small population sizes, which is in itself surprising given its state space is many orders of magnitude greater than can be explored in any realistic or relevant evolutionary timescale \cite{mcleish:2015}. This ergodic property is closely related to the idea of space-shape covering, where the property of high dimensional maps means many phenotypes are potentially accessible from each other by only a few mutations. The main result is that in this more complex GP map, reproductive isolation also arises more quickly for small populations \cite{khatri:2019}, which is related to the strong phenotypic bias. In addition, analogous to transcription factor DNA binding, it is also found that the molecular binding energy phenotypes---that underlie the organismal level patterning phenotype---which are under weakest selection, are most likely to give rise to the earliest hybrid incompatibilities.

Altogether, these results point to a universal picture to understand divergence between populations and the role of population size for strongly conserved traits; high-fitness phenotypes tend to be also highly specified, which means in converse low fitness phenotypes will have a large relative degeneracy or phenotypic bias. This means that the balance between fitness and sequence entropy, embodied by the maximum of free fitness, will be a strong feature of the equilibrium probability distribution of strongly conserved phenotypes, which are under stabilising selection. For simple biophysical traits like transcription factor DNA binding or protein stability, it is clear this is true since there will always only be a few sequences that give maximum affinity or stability, however, this has been found to be true even in a more complex GP map for developmental system drift. It is likely that in some way the sequence entropy constraints of transcription factor DNA binding propagate up in determining the sequence entropy of the organismal level patterning phenotype. The open questions are: how universal is this phenomenon?, will far more complex GP maps also show this behaviour?, will such maps maintain their property of ergodicity?, and is there a broad theoretical framework that can address this question without the more complex and computationally intensive simulations needed to address the former? Beyond an equilibrium analysis, there is the open question of dynamics and adaptation in GP maps \cite{manrubia:2015,khatri:2015,schaper:2014,nourmohammad:2013}, as well as extending this formalism to the strong mutation regime, yet still at finite population size (as compared to the infinite population size, quasispecies regime \cite{iwasa:1988,barton:2009,nourmohammad:2013,khatri:2018}).

\section{GP maps as evolving objects}\label{sec:evolutionOFgpmaps}

Important new insights on quantitative features of adaptation have been obtained by studying evolutionary processes with realistic, highly nonlinear GP maps, as presented in previous sections. However, the concept of a predefined GP map on which evolutionary processes occur is not realistic. Not only should we expect the phenotype-to-fitness relation to vary due to environmental fluctuations---changing the fitness landscape into a seascape \cite{mustonen:2009}---, but it is the GP map itself that is subject to evolution. Indeed, one might argue that in the long term what occurs is the evolution \emph{of} the GP map, rather than a simple adaptation on a sort of preexisting genotype space.

By evolution of the GP map we mean two things: first, that the assignment of phenotypes to genotypes is a dynamic process that depends on context. As a consequence, the same genotype can present very different phenotypes during the course of evolution. And, second, that the dimensionality of the map changes during the course of evolution \cite{zeldovich:2007_PLoSCB}. Indeed, duplications, deletions or large-scale chromosomal rearrangements, among others, are very frequent and often related to the acquisition of new or different phenotypic features \cite{kent:2003}.

In the following we will explore two computational models
where the GP map itself is allowed to evolve, each demonstrating one of the features of GP map evolution mentioned above. We focus our discussion on the evolution of mutational neighbourhood, that determines which phenotypes are accessible from an evolved (evolving) genotype. As we will see, evolved populations in these two models fine-tune their mutational neighbourhoods so that adaptive phenotypes arise more frequently as a result of mutation. It appears that the explicit consideration of the mutational neighbourhood determined by the evolution of the GP map is essential for understanding not only long and short term evolution, but also the functioning of present day organisms.

\subsection{Evolution of a multifunctional quasispecies in an RNA world model}
\label{sec:Quasispecies}

\begin{figure*}
\centering
\includegraphics[width=1.0\textwidth]{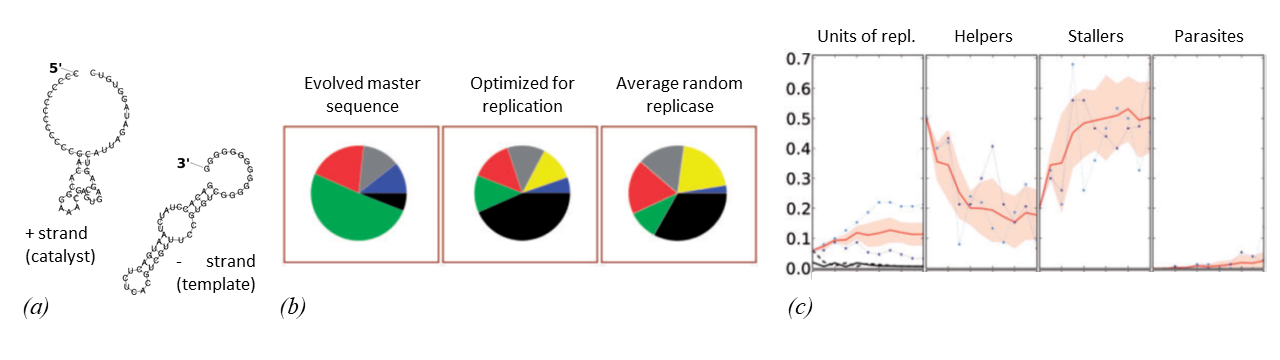}
\caption{The region of the genotype space selected by a population within an RNA world model is highly special as compared to controls. (a) Secondary structure of a replicase ($+$ strand) and its $-$ strand which is optimized for maximum replication rate. (b)  Functional classes in the 1-mutation-away mutational neighbourhood. Black: replicases; yellow: parasites; green: helpers; red: stallers; gray: junk; blue: unclassified. From left to right, pie charts correspond to a purely evolved replicase, a replicase  optimized for maximum replication rate, and an average random replicase, respectively. Strong reduction of replicases and parasites and strong over-representation of helpers convey robustness to high mutation rates. (c) 1- to 10-mutations-away neighbourhoods ($x$-axis of each functional type) of the evolved GP map: at larger mutational (and therewith spatial) distances, frequencies of helper mutants decline drastically, whereas the frequency of stallers increases,  thereby preferentially helping the ancestor and stalling others, including parasites.} 
\label{fig:evolvedRNAmap}
\end {figure*}

The RNA World model \cite{gilbert:1986} envisages a plausible scenario for the origin and early evolution of life. Understanding how the RNA World could have arisen involves explaining how diverse molecular function might emerge in the absence of faithful replication. Interestingly, it has been suggested that phenotypic bias could have played a main role in solving this problem \cite{briones:2009}. Evolution and selection become possible only once the replication machinery is in place. In perspective, two alternative approaches have been extensively used for studying evolution of the RNA world: those that study the evolution on the RNA-sequence-to-secondary-structure GP map and those studying the impact of spatial pattern formation on what is selected. While the former class of models study the RNA world using the GP map with predefined fitness criteria, the latter explores the eco-evolutionary dynamics of replicator interactions without a predefined fitness. These two approaches have been combined \cite{takeuchi:2008,colizzi:2014} in a case study of the evolution of the qualitative, emergent functional properties of mutational neighbourhoods. 

In this model, RNA sequences are embedded in a 2D grid and interact with their closest neighbours by complementary base pairing, forming complexes. If one of the molecules $X$ folds into a structure with pre-dened motifs and binds to a molecule $Y$, replication can occur and the complementary strand of the molecule $Y$ is formed. No fitness is explicitly defined, and therefore it arises as an emergent property of the population. Because of the spatial embedding, the interactions that occur are shaped by emergent spatial structures. Such emergent spatial structures constitute a new level of selection and deeply affect the evolutionary outcome of replicators (as it has been shown in previous examples related to the RNA world \cite{boerlijst:1991,takeuchi:2009}).  

At all mutation rates studied, replicases rapidly evolve symmetry breaking between the complementary RNA strands, with one strand having replicase functionality and the complementary strand evolving an optimal template function---i.e. optimally binding the replicase. This symmetry breaking is also seen in non-GP-map-based toy models \cite{takeuchi:2017,dunk:2017}. The stationary phenotypic composition of the population, however, strongly depends on mutation rate. At high mutation rates, only one, highly polymorphic quasispecies of replicases exists, whereas at lower mutations rates multiple quasispecies coexist. At intermediate mutation rates, there is coexistence between replicases and parasites, RNA molecules that act as templates for the replicases but which have no catalytic function themselves. At lower mutation rates, two different replicase-parasite communities coexist. Finally, at the lowest mutation rates these communities compete with each other sometimes going to extinction.

\begin{figure}
\begin{infobox}[frametitle= Box~\ref{box:mutationalclasses}: Emergent functional classes in an RNA world model]
{In the RNA world model here described \cite{takeuchi:2008,colizzi:2014}, molecular phenotypes were determined for pairs of complementary sequences ($+$ and $-$
  strands), based on specific structure motifs. Under evolution, the following phenotypes emerge: \\
  {\bf Self-replicases} can replicate both other molecules as well as themselves. \\
  {\bf Parasites} are RNA sequences that only work as templates and have no replicase ability. \\
  {\bf Helpers} can  replicate  other  molecules  but cannot be replicated. \\  
  {\bf Stallers} can engage molecules in complexes, but can neither replicate them nor be replicated. \\  
  {\bf Junk} cannot form complexes and are therefore mostly inert.
}
\label{box:mutationalclasses}
\end{infobox}
\end{figure}

Let us focus now on the mutational neighbourhood of the replicases that evolve at the highest sustainable mutation rates. The functional composition (see Box~\ref{box:mutationalclasses}) of the mutational neighbourhood of such evolved replicases is compared to two controls, i.e. a replicase that has been optimised for its replication rate, and randomly sampled replicases (Fig.~\ref{fig:evolvedRNAmap}b). In the mutational neighbourhood of the evolved replicases, replicases are scarce, parasites are missing, helpers are over-represented, and non-viable stallers are above average. In contrast, the controls have many replicases and parasites, and much fewer helpers (Fig.~\ref{fig:evolvedRNAmap}b).  

The advantages of the multifunctional organisation of the mutational neighbourhood can be understood as follows. Non-viable mutants tend to be spatially close to their ancestor. Thus the helpers, in the close mutational neighbourhood, tend to help their ancestor and siblings rather than others. The non-viable helpers are essential for survival: if they are eliminated, the whole system goes extinct. This advantage of helpers is true only because there are no parasites in the mutational---and therefore spatial---neighbourhood. In contrast, stallers are detrimental for the system, but less so for their ancestor, for whose survival they are essential. This is because there are fewer stallers in the close neighbourhood than farther away (Fig.~\ref{fig:evolvedRNAmap}b and c) and they therefore hinder others more than the ancestor. In particular, they stall parasites if they emerge farther away. Indeed, if stallers are killed, parasites invade the system forming the two-species system characteristic of lower mutation rates. 

In this scenario, functions were not pre-conceived, but
emerged. Because the implemented GP map is actually the classical RNA GP map, and only point mutations are considered, the evolutionary dynamics could be seen as evolution \emph{on} this fixed GP map. However, the GP map described in terms of these functions, and their structural implementation, fits better in the conceptualisation of evolution \emph{of} the GP map, where some phenotypes (and thus functions, like helpers or stallers) evolve not as separate lineages, but as mutants in the evolved mutational neighbourhood of a replicase.

\subsection{Evolution of genome size and evolvability in virtual cells}
\label{sec:VirtualCells}

\begin{figure*}
\centering
\includegraphics[width=1.0\textwidth]{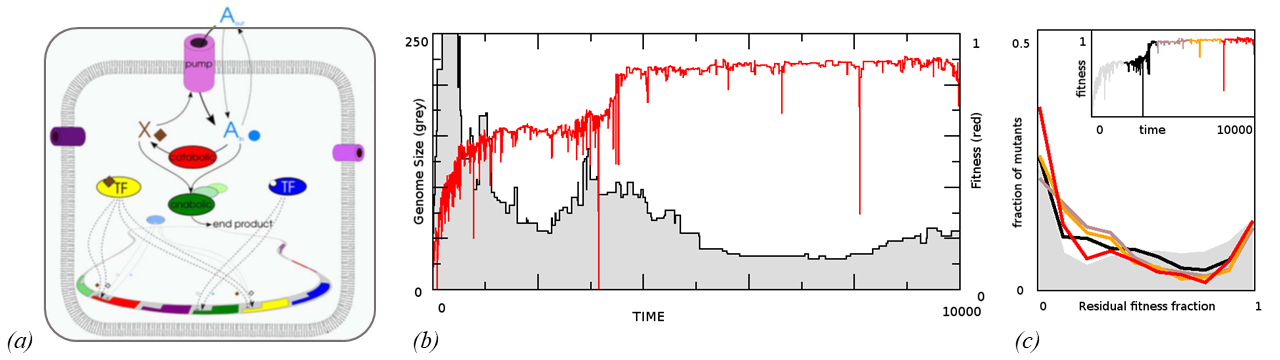}
\caption {Virtual cell and evolutionary dynamics. (a) Scheme of a virtual cell. (b) Common ancestor through time. Red line is the average fitness in three standard environments; shaded grey area depicts genome size, showing initial genome inflation followed by streamlining (gene loss). (c) Mutational neighbourhood of the ancestor in various time periods, colour coded according to inset: during evolution the mutational neighbourhood changes from the initial fitness distribution of the shaded grey area to a more pronounced ``U-shape'', with peaks at neutral mutations (right side) and strongly deleterious mutations (left side).}
\label{fig:virtualcell}
\end {figure*}

Now we explore {\it virtual cells} (Fig.~\ref{fig:virtualcell}a), a second model where we allow the dimensionality of the GP map to change while fixing a fitness function \cite{cuypers:2012,cuypers:2014,cuypers:2017}. The system consists of a genome with genes coding for enzymes, pumps and transcription factors, as well as transcription factor binding sites. The transcribed genes form a simple metabolic network, which pumps in resources and transforms them into energy and building blocks. The external resource fluctuates, and fitness is homeostasis: the energy carrier and internal resource have to be close to a preset value. Average homeostasis over a cell's lifetime determines its fitness at replication. Mutations include changes in parameter values as well as gene duplications, deletions and large chromosomal rearrangements. Thus, genome size is variable. 

Figure~\ref{fig:virtualcell}b summarises the dynamics of evolved virtual cells at different stages. Early in evolution, genome size expands dramatically and immediately declines sharply. Although this transient is not generic, it occurs in those evolutionary runs which later reach high fitness. Interestingly, genome expansion does not entail an immediate fitness benefit, since there is no difference, in this time frame, between runs in which the common ancestor does and does not expand its genome. Subsequently, gene loss dominates the evolutionary dynamics most of the time, and often conveys increases in fitness. 

The mutational neighbourhood, represented here as the fraction of mutants with decreasing fitness, has a characteristic ``U-shape'' (Fig.~\ref{fig:virtualcell}c), which becomes more pronounced during evolution. The fraction of neutral mutants remains the same despite an overall fitness increase, whereas the fraction of lethal mutations increases and the fraction of slightly deleterious mutations decreases. 

Through this process, populations become highly evolvable. After a drastic environmental change (here implemented as a change in basic parameters) it sometimes takes only a few minor mutations to recover from nearly zero fitness to a value comparable to that previous to the environmental change; in other cases, a relatively fast recovery of fitness is mediated by genome expansion. After repeated environmental switches, evolvability through few mutations becomes common. Such fast evolvability turns out to be easier to evolve than regulatory mechanisms to adapt to changing environments \cite{cuypers:2017}.

These results are consistent with experimental reports. Phylogenetic reconstructions of long-term evolution show surprisingly large genome sizes of common ancestors (LUCA and LECA) and evolutionary dynamics dominated by gene loss \cite{koonin:2007}. The U-shape mutational neighbourhood has been observed in yeast \cite{wloch:2001} and viruses \cite{sanjuan:2004}. Lastly, fast adaptation to environmental changes mediated by few mutations or by genome expansion are well documented in many evolutionary experiments, for instance in yeast \cite{yona:2012}. Antibiotic production in {\it Streptomyces} is done by highly unfit mutants \cite{zhang:2019}, an evolutionary signature resembling the multifunctional quasispecies described in the RNA example. It is remarkable that all these surprising evolutionary signatures emerge in a minimal cell model, suggesting that they are generic features of Darwinian evolution, if genome organisation and the GP map are allowed to evolve.

\section{Empirical genotype-to-phenotype and genotype-to-function maps}
\label{sec:empirical}

Technological advances are facilitating the experimental characterization of GP maps at ever-increasing resolution and scale \cite{devisser:2014,payne:2019}. The phenotypes of such maps include the activity or binding specificity of macromolecules such as RNA and proteins \cite{olson:2014,pitt:2010,sarkisyan:2016,diss:2018}, the exonic composition of transcripts \cite{julien:2016}, the spatiotemporal gene expression pattern of regulatory circuits \cite{schaerli:2014,schaerli:2018}, as well as the function and flux of metabolic pathways \cite{bassalo:2018}. In some cases, it is even possible to measure organismal fitness {\it en masse} \cite{li:2016,puchta:2016,venkataram:2016,rotem:2018}.

When combined with a mapping from phenotype to fitness, biophysical GP maps provide a principled approach to constructing a fitness landscape over the space of genotypes. In situations where an empirical genotype-fitness map is available but a mechanistic understanding of its structure is lacking, one may instead try to infer the hidden phenotypic level from the genotype-fitness data. Ideally the inferred phenotypes can be interpreted biologically, but even when this is not the case, the introduction of an intermediate phenotypic layer helps to organise the high-dimensional genotypic data set and to reduce its complexity. 

In this section, we begin with GP maps that have been empirically characterised. Sometimes, the quantity that is experimentally accessible is fitness, and not phenotype. We discuss how empirical data of that kind can be used to infer the structure of fitness landscapes and some properties of the phenotypic level. Then, we delve into the characterisation of GP and genotype-to-function maps in virus populations, discussing as well the implications of those maps in evolutionary dynamics under constant and variable environments. Next we address the inference of intermediate phenotypes from genotype-fitness data and, finally, we discuss approaches to the experimental characterisation of GP maps that may be relevant to synthetic biologists.

\subsection{Empirical GP maps}

There are three main approaches for constructing empirical GP maps \cite{devisser:2014,payne:2019}: (i) a combinatorially complete map is constructed using all possible combinations of a small set of mutations, such as those that occurred along an adaptive trajectory in a laboratory evolution experiment or in natural history \cite{weinreich:2006}; (ii) a deep mutational scan assays the phenotypes of all single mutants, as well as many double- and triple-mutants of a single wild-type genotype \cite{fowler:2014}; and (iii) an exhaustively-enumerated map is constructed from all possible genotypes---something which is only possible for very small genotype spaces \cite{rowe:2009,jimenez:2013,payne:2014b}. In some cases, such as with antibody repertoires \cite{adams:2019,miho:2019} or viral populations \cite{hinkley:2011,acevedo:2014}, a fourth method of construction is possible. Specifically, one can directly construct a small portion of an empirical GP map by collecting a large number of genotypes with a particular phenotype from nature (e.g., the ability of an antibody to bind an antigen). Below, we describe a recent example from each of the three main categories, highlighting the biological insights gained from the construction and analysis of such maps.

\subsubsection{A combinatorially complete map} 

Alternative splicing is a key step of post-transcriptional gene regulation, and exonic mutations that affect splicing are commonly implicated in disease \cite{daguenet:2015}. All possible combinations of mutations that occurred in the evolution of exon 6 in the human {\it FAS} gene since the last common ancestor of humans and lemurs have been analysed \cite{baeza:2019}. A total of 3,072 genotypes were assayed for the percentage of transcript isoforms that included the exon. This phenotype of ``percentage spliced-in" varied from 0\% to 100\% among the 3,072 genotypes, indicating that in combination, these mutations are capable of producing the full range of exon inclusion levels. Importantly, the phenotypic change induced by a mutation depended non-monotonically upon the phenotype of the genotype in which the mutation was introduced, such that mutations to genotypes near the full-exclusion or full-inclusion phenotypic bounds had the smallest effects, whereas mutations to genotypes with intermediate inclusion levels had the largest effects. The resulting biological insight is that the evolution of an alternative exon from a constitutive exon will require several mutations, because mutation effect sizes are smallest when the exon is near full inclusion. This observation led to the mathematical derivation of a scaling law that applies to this and possibly other GP maps, and that may aid in the development of drugs aimed at targeting splicing for therapeutic benefit, by helping to predict drug-sensitive splicing events.

Another example of a combinatorially complete map will be explored more fully in Section \ref{sec:viruslandscape}.

\subsubsection{A deep mutational scan assay}

Amino acid metabolism is fundamental to life, and is driven by complex metabolic and regulatory pathways. A deep mutational scan of nineteen genes involved in four pathways that affect lysine flux in {\it E. coli} was performed \cite{bassalo:2018}. The resulting GP map consisted of 16,300 genotypes, each of which was assayed for its resistance to a lysine analogue that induces protein misfolding and reduces cell growth. The phenotype was therefore grown in the presence of the analogue. Several resistance-conferring mutations were identified, including mutations in transporters, regulators, and biosynthetic genes. For example, such mutations were often observed in a lysine transporter called LysP. These were relatively evenly distributed across the gene, suggesting that loss-of-function mutations were a common evolutionary path toward abrogated transport of the lysine analogue. More generally, this study represents a proof-of-concept that deep mutational scanning experiments can be scaled up from individual macromolecules to regulatory and metabolic pathways.

\subsubsection{An exhaustively enumerated map} 
\label{sec:EEM}

Binding of regulatory proteins to DNA and RNA molecules are central to transcriptional and post-transcriptional gene regulation, respectively. The robustness and evolvability of these two layers of gene regulation has been studied via a comparative analysis of two empirical GP maps \cite{payne:2018}. At the transcriptional level, interactions between DNA and transcription factors were considered, where a genotype was a short DNA sequence (a transcription factor binding site) whose phenotype was its molecular capacity to bind a transcription factor. At the post-transcriptional level, interactions between RNA and RNA binding proteins were analysed, where a genotype was a short RNA sequence (an RNA binding protein binding site) whose phenotype was the capacity to bind an RNA-binding protein. Though robustness at both layers of gene regulation was comparable, there were marked differences in evolvability, which were suggestive of qualitatively different architectural features in the two GP maps. Specifically, the genotype networks of binding sites for RNA binding proteins were separated by more mutations than the genotype networks of binding sites for transcription factors, rendering mutations to the binding sites of RNA binding proteins less likely to bring forth phenotypic variation than mutations to the binding sites of transcription factors. These observations are consistent with the rapid turnover of transcription factor binding sites among closely related species, as well as with the relatively high conservation levels of binding sites for RNA binding proteins. This comparative analysis may therefore help to explain why transcriptional regulation is more commonly implicated in evolutionary adaptations and innovations than post-transcriptional regulation mediated by RNA binding proteins.

\subsection{Empirical fitness landscapes and adaptive dynamics of viral populations}\label{sec:viruslandscape}

\begin{figure}
\begin{infobox}[frametitle= Box~\ref{box:epistasis}: Epistasis and fitness landscapes]
{
\begin{center}
    \includegraphics[width=0.5\textwidth]{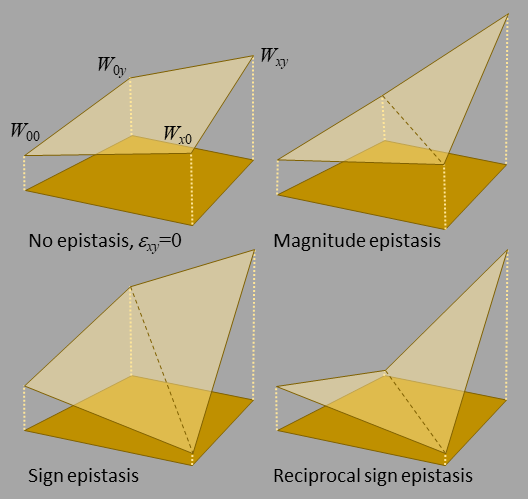}
\end{center}
    \label{fig:epistasis}
Epistasis means that the phenotypic effects of a mutation depend on the genetic background (genetic sequence) in which it occurs \cite{poelwijk:2007}. Whereas the concept applies mainly to any phenotypic trait, in the evolutionary context epistasis for fitness is of primary importance, and we will focus on this case in what follows. The degree of epistasis, $\epsilon_{xy}$, between a pair of mutations $x$ and $y$ can be estimated as $\epsilon_{xy} = W_{00}W_{xy} - W_{x0}W_{0y}$, where $W_{00}$ is the fitness of the non-mutated genotype, $W_{xy}$ the experimentally determined fitness of the double mutant and $W_{x0}$ and $W_{0y}$ are the measured fitness of each single mutant. Under a multiplicative fitness effect model, $W_{x0}W_{0y}/W_{00}$ represents the expected fitness value of the double mutant and, therefore, $\epsilon_{xy}$ represents the deviation from this null hypothesis. The sign of $\epsilon_{xy}$ corresponds to the sign of epistasis. \\[2mm]
Magnitude epistasis causes deviations from the multiplicative model, but the landscape remains monotonic; sign epistasis means that the fitness sign of at least one of the mutations in the pair changes in presence of the other mutation; reciprocal sign epistasis occurs when both mutations change the sign of their fitness effect when combined, so both potential adaptive pathways connecting the nonmutated ancestor with the double mutant necessarily must cross a valley. Epistasis thus determines the ruggedness of a fitness landscape \cite{wright:1932,weinreich:2005,poelwijk:2011} and therefore the accessibility of adaptive pathways \cite{schaper:2011}. If there is either magnitude epistasis or no epistasis at all, fitness landscapes are smooth and single-peaked, and evolving populations can reach the global maximum. In the case of sign epistasis, only a fraction of the total paths to the optimum are accessible. Reciprocal sign epistasis is a necessary but not sufficient condition for rugged landscapes with multiple local optima \cite{poelwijk:2011}, a situation where an evolving population might get stuck into suboptimal peaks. Most studies on epistasis have focused on pairwise epistasis, ignoring interactions among more than two mutations. However, higher-order epistasis appears in almost every published combinatorial fitness landscape \cite{weinreich:2013:COGD}, so the topographical features of fitness landscapes seem to depend on all orders of epistasis.
}
\label{box:epistasis}
\end{infobox}
\end{figure}

 The topography of fitness landscapes is key to understand evolutionary dynamics, and recent studies have focused on epistasis as a measure of landscape ruggedness (see Box~\ref{box:epistasis}). Two different experimental approaches have been taken to characterise the ruggedness of fitness landscapes through epistasis: a first, simpler approach is to analyse the epistasis among random pairs of mutations \cite{elena:1997,bonhoeffer:2004,sanjuan:2004,lalic:2013}, while a more exhaustive approach relies on reconstructing a combinatorial fitness landscape that includes all possible combinations among a set of $m$ mutations \cite{devisser:2014}. Usually, these $m$ mutations have been observed during experimental evolution and adaptation of populations to novel environments. Such empirical landscapes have been characterised for bacteria \cite{lunzer:2005,weinreich:2006,poelwijk:2007,dawid:2010,chou:2011,khan:2011}, protozoa \cite{lozovsky:2009}, fungi \cite{devisser:2009,hall:2010}, and human immunodeficiency virus type-1 (HIV-1) \cite{dasilva:2010,hinkley:2011,kouyos:2012,dasilva:2014}.

In what follows, we review work focusing on the topography of an RNA virus fitness landscape. We begin with an investigation of how prevalent different epistasis types are (see Box~\ref{box:epistasis}), and then continue with the influence of landscape topography on the evolutionary potential of a virus population. Finally, we discuss the relevance of the environment on viral evolution, through analyses of landscapes on different host species.

\subsubsection{Description of epistasis among random pairs of mutations}

The analysis of the effects of mutations on fitness provides information about the degree of ruggedness of the landscape at a coarse-grained level. In a study with {\it Tobacco etch potyvirus} (TEV) \cite{lalic:2012}, 20 single nucleotide substitution mutations randomly scattered along the RNA genome of the virus were analysed. These mutations were deleterious when evaluated in {\it Nicotiana tabacum,} its natural host, through competition experiments against a reference TEV strain \cite{carrasco:2007}. Those single mutations were randomly combined to yield 53 double mutants, whose fitness was measured also in {\it N.~tabacum}. Twenty combinations rendered $\epsilon_{xy}$ values significantly deviating from the null expectation, 11 of which were positive and 9 negative (see Box.~\ref{box:epistasis}). Interestingly, these nine cases were all examples of synthetic lethality, that is, single mutations were deleterious but viable, but in combination became lethal. This represents an extreme case of negative epistasis.  
Previous studies with other RNA viruses obtained comparable epistatic interactions in type and sign \cite{bonhoeffer:2004,sanjuan:2004,sanjuan:2006:JGV}.

How can we explain positive epistasis in the small and compact genomes of RNA viruses? Given the lack of genetic and functional redundancy and, in many cases, overlapping genes and multifunctional proteins, a small number of mutations can produce a strong deleterious effect. But, as mutations accumulate, they affect the same function with increasing probability and thus, their marginal contribution to fitness diminishes. Hence, the observed fitness is above the expected multiplicative value. In other words, epistasis is positive.

\subsubsection{Description of a combinatorial landscape and higher-order epistasis}

\begin{figure*}[ht]
\begin{center}
\includegraphics[width=1.0\textwidth]{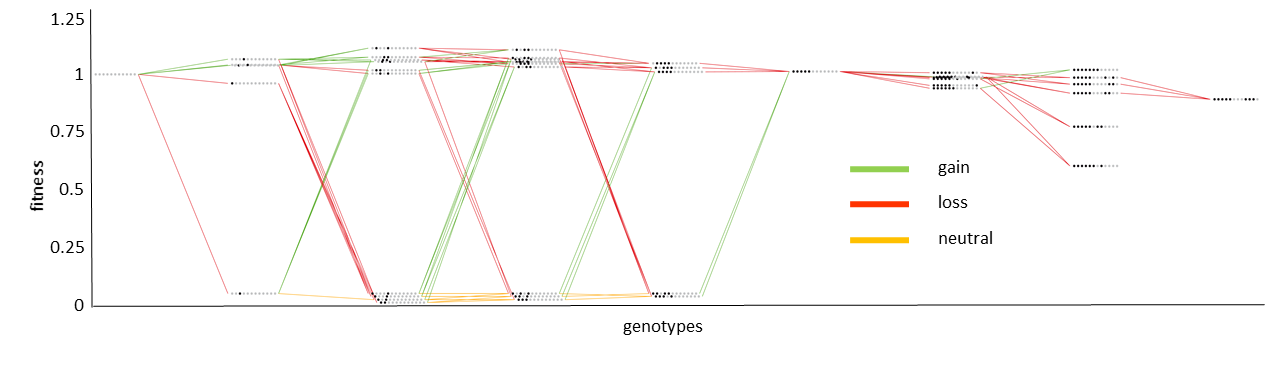}
\end{center}
\caption{Snapshot of an empirical fitness landscape constructed with combinations of mutations observed during experimental adaptation of tobacco etch potyvirus (TEV) to its new experimental host {\it Arabidopsis thaliana}. Each string of dots represents a genotype. Black dots represent a mutation in the corresponding locus, while grey dots correspond to the wild-type allele at that locus. Green lines stand for mutations with beneficial effect, red lines for deleterious mutations and orange lines for neutral mutations in the corresponding genetic background. Lines link genotypes which are one mutation away. The global optimum for this landscape corresponds to the 01001000 genotype. Data has been processed with MAGELLAN, which qualitatively orders genotypes along the $x-$axis according to the number of mutations. \cite{brouillet:2015}}
\label{fig:LandscapeTEV}
\end{figure*} 

TEV was evolved in a novel host, {\it Arabidopsis thaliana}, until it achieved high fitness \cite{agudelo-romero:2008}. The consensus genome of this adapted strain had only six mutations, three of which were nonsynonymous. The fitness effect of five of these mutations (the sixth one had to be discarded) was individually evaluated: two were significantly beneficial (one synonymous and one nonsynonymous), one was neutral (nonsynonymous), one deleterious (synonymous), and one lethal (nonsynonymous) \cite{agudelo-romero:2008}. All $2^5 = 32$ possible genotypes that result from combining the observed five mutations were created in order to generate a complete five-sites landscape (Fig.~\ref{fig:LandscapeTEV}), with abundant epistasis. Thus, the obtained landscape was rugged and without neutrality.

The pervasiveness of higher-order epistatic interactions in all empirically characterised combinatorial landscapes \cite{weinreich:2013:COGD} prompted its study in the small TEV combinatorial landscape \cite{lalic:2015}. Using the Walsh-transform method \cite{weinreich:2013:COGD,poelwijk:2016}, higher-order epistatic interactions were found to be as important as pairwise interactions to fully understand the topological properties of adaptive landscapes.

Interestingly, and despite previous reports claiming that pervasive epistasis results in predictable evolutionary dynamics \cite{devisser:2014}, repeated evolutionary experiments starting from different genotypes of the TEV virus resulted in different evolutionary endpoints, and populations were able to escape local optima, moving efficiently in this highly rugged landscape, with new mutations appearing in the course of evolution. \cite{cervera:2016:PRSB} This result suggests that evolutionary predictions based on extrapolations from non-exhaustive fitness landscapes have to be taken with care, as evolving populations are often able to find new, previously undescribed mutations that introduce new evolutionary dimensions. 

\subsubsection{Effect of host species on the topography}

Another quite common observation in evolutionary experiments with RNA viruses, as well as in natural populations, is the existence of pleiotropic fitness costs across different hosts \cite{bedhomme:2015}---beneficial mutational effects in one host may become deleterious in an alternative host. These negative fitness effects limit the host range of viruses to closely related species that share most of the molecular targets needed for the virus to complete its infectious cycle.

The concept of pleiotropy can be explored in terms of changes in the topography of fitness landscapes across hosts. The fitness of TEV single and double mutants was measured in four different susceptible hosts that differed in their degree of genetic relatedness \cite{lalic:2013}: the natural host {\it N.~tabacum, Datura stramonium} (in the same botanical family, Solanaceae), {\it Helianthus annuus} (an Asteraceae phylogenetically related to the Solanaceae---both are Asterids), and in {\it Spinacea oleraceae} (an Amaranthaceae).  Both the sign and the magnitude of epistasis changed across hosts: epistasis was positive ($\epsilon_{xy} > 0$) only in the natural host, and it diminished as the host species relatedness to {\it N.~tabacum} decreased.

The topography of the combinatorial fitness landscape was more rugged in {\it N.~tabacum} when evaluated with the TEV strain adapted to {\it A.~thaliana} \cite{cervera:2016:JV}. Though the global optimum was the same in both landscapes, it was less accessible in {\it N.~tabacum} given the greater magnitude of reciprocal sign epistasis in its vicinity.

Altogether, these results suggest that the topography of the adaptive fitness landscape for an RNA virus is strongly dependent on the environment (host species), though some general properties, such as the existence of lethal genotypes, minimal or null neutrality and high ruggedness, remain. In this light, novel frameworks that explicitly account for environmental changes on the properties of landscapes, such as seascapes or adaptive multiscapes \cite{catalan:2017}, should yield a better picture to think about these experiments and even provide some predictive power.

\subsection{Inferring phenotypes from genotype-fitness maps}
\label{inferring-phenots}

Massively parallel empirical studies that examine a large ensemble of genotypes often yield information on their biological activity, or their overall fitness, while the identification of the phenotypes involved becomes difficult. In such cases it is possible to infer a phenotypic level, ideally endowed with a biological meaning, from data analogous to that of the previous sections. The key assumption underlying these formal approaches is that the mutational effects on the unobserved phenotypic traits are additive, such that any epistatic interaction for fitness arises from the nonlinearity of the phenotype-fitness map \cite{domingo:2019b}. In the simplest case of a one-dimensional trait that maps monotonically to fitness, the trait variable has been referred to as the \textit{fitness potential} \cite{kondrashov:2001,milkman:1978} and the nonlinearity of the phenotype-fitness map as \textit{global epistasis} \cite{otwinowski:2018}. Because a monotonic phenotype-fitness map preserves the rank ordering of genotypes with respect to fitness, it can account for magnitude epistasis, but not for sign epistasis \cite{weinreich:2005}.

Sign epistasis can however arise from non-monotonic one-dimensional phenotype-fitness maps. Consider a fitness function $f(x)$ where the wild type trait value is located at $x=0$ and grows toward a single phenotypic optimum at $x_\mathrm{opt} > 0$. A mutation that increases the trait value by an amount $\Delta x < x_\mathrm{opt}$ is then beneficial on the wild type background but deleterious on a background with trait value $x \geq x_\mathrm{opt}$ that overshoots the optimum. In an experimental study of the ssDNA bacteriophage ID11, it was found that this scenario explains the pairwise epistatic interactions between 9 individually beneficial mutations rather well \cite{rokyta:2011}. In this case, the phenotype-fitness map was taken to be a gamma function with 4 parameters, and the unknown phenotype was parametrised by the 9 single mutational effects. The joint inference of the phenotypic effects and the phenotype-fitness map thus required 13 parameters to be estimated from the fitness values of 9 single and 18 double mutants. 

The range of epistatic interaction patterns that can be generated from a one-dimensional phenotypic trait subject to a single-peaked phenotype-fitness map is obviously limited. In particular, any evolutionary trajectory composed of mutations that are individually beneficial on the wild-type background can display at most one fitness maximum. This criterion was used in a recent study of the combined resistance effects of synonymous mutations in the antibiotic resistance enzyme TEM-1 $\beta$-lactamase challenged by cefotaxime to conclude that the phenotype underlying these effects is most likely multidimensional \cite{zwart:2018}. Multidimensional phenotypes allow for more versatile interaction structures but also require more parameters to be inferred from data. In a study of non-synonymous resistance mutations in TEM-1, a two-dimensional phenotype combined with a sigmoidal phenotype-fitness map was found to provide a good description of the measured resistance values \cite{schenk:2013}. In this case one of the phenotypes was taken to be protein stability, which was determined computationally, whereas the second phenotype was inferred along the lines of the experiment with the ID11 bacteriophage above \cite{rokyta:2011}. A similar approach has been applied to the fitness landscape of a norovirus escaping a neutralising antibody, where the folding stability and binding affinity of the capsid protein were mapped to the probability of infection \cite{rotem:2018}. Importantly, in two-dimensional phenotype-fitness maps sign epistasis can emerge even in the absence of a phenotypic optimum \cite{manhart:2015,schenk:2013}. 

\begin{figure}[ht]
\begin{center}
\includegraphics[width=8cm]{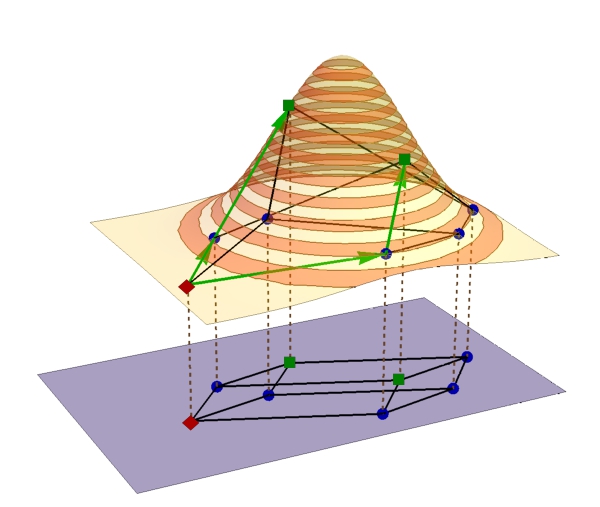}
\end{center}
\caption{\label{JK-Fig1} Illustration of Fisher's geometric model for a two-dimensional phenotype and a single-peaked phenotype-fitness map. Three phenotypic mutations originating from the wild type (marked in red) combine additively, giving rise to a distorted three-dimensional cube in the phenotype plane. As a consequence of the nonlinear mapping to fitness, two of the double mutants (marked in green) become local fitness maxima in the induced genotypic landscape. Courtesy of Sungmin Hwang.}
\end{figure}

The models described so far can be viewed as variants of the \textit{geometric model} devised by Ronald Fisher to argue that the adaptation of complex phenotypes must proceed in small steps \cite{blanquart:2014,fisher:1930,tenaillon:2014}. Originally, Fisher's geometric model (FGM) did not include the assumption of additive phenotypes, which was introduced later in a study of pairwise epistasis between mutations in \textit{Escherichia coli} and vesicular stomatitis virus \cite{martin:2007}. In its modern formulation, the model is based on a set of $d$ real-valued traits forming a vector $\vec{x} = (x_1,x_2,\cdots,x_d)$ in the $d$-dimensional Euclidean space $\mathbb{R}^d$ and a nonlinear phenotype-fitness function $f(\vec{x})$ with a single optimum that is conventionally located at the origin $\vec{x}=0$ (Fig.~\ref{JK-Fig1}). The genotype is described by a sequence $\tau = (\tau_1, \tau_2,\cdots,\tau_L)$ of $L$ symbols $\tau_i$ drawn from the allele set $\{0,1,\cdots,k-1\}$, where $\tau_i = 0$ denotes the wild type allele and in most cases a binary alphabet with $k=2$ has been considered. The additive GP map takes the form \cite{hwang:2017}
\begin{equation}
\label{JK:FGM}
\vec{x}(\tau) =  \vec{x}_0 + \sum_{a=1}^{k-1} \sum_{i=1}^L \delta_{\tau_i,a} \vec{v}_{i,a},
\end{equation}    
where $\vec{x}_0$ is the wild type phenotype and the vector $\vec{v}_{i,a} \in \mathbb{R}^d$ describes the phenotypic effect of the mutation $0 \to a$ at the $i$'th genetic locus. The genotype-fitness map is then obtained as $F(\tau) = f[\vec{x}(\tau)]$. In applications of FGM to experimental data, the mutational effects $\vec{v}_{i,a}$ are usually treated as random vectors drawn from a multivariate Gaussian distribution. Rather than inferring specific phenotypes, such analyses yield gross statistical features of the phenotypic landscape, such as the number of phenotypic traits $d$, the distance of the wild type phenotype from the optimum $\vert \vec{x_0} \vert$, and the variance of phenotypic mutational effects \cite{blanquart:2016,martin:2007,schoustra:2016,weinreich:2013:E}.

Current high-throughput sequencing methods are capable of measuring fitness and other functional phenotypes for hundreds of thousands of genotype sequences in a single experiment, and methods based on the inference of unobserved additive traits provide an important tool for organising and interpreting the resulting data sets. A recent large-scale analysis of the fitness landscape of a segment of the \emph{His3} gene in yeast built out of amino acid substitutions from extant species made use of a deep learning approach to infer the additive phenotype and its nonlinear mapping to fitness \cite{pokusaeva:2019}. Remarkably, large parts of the data were well described by a one-dimensional fitness potential combined with a sigmoidal phenotype-fitness map, suggesting that much of the observed complexity of epistatic interactions could potentially be explained in terms of thermodynamic considerations \cite{otwinowski:2018}. Combining such data-driven inference methods with biophysical modeling and functional information appears to be a promising route towards a deeper understanding of the relation between genotype, phenotype and fitness on the molecular level \cite{bershtein:2017}. 

\subsection{Synthetic biology approaches to characterising GP maps}

\begin{figure*}[t]
\centering
\includegraphics[width=\textwidth]{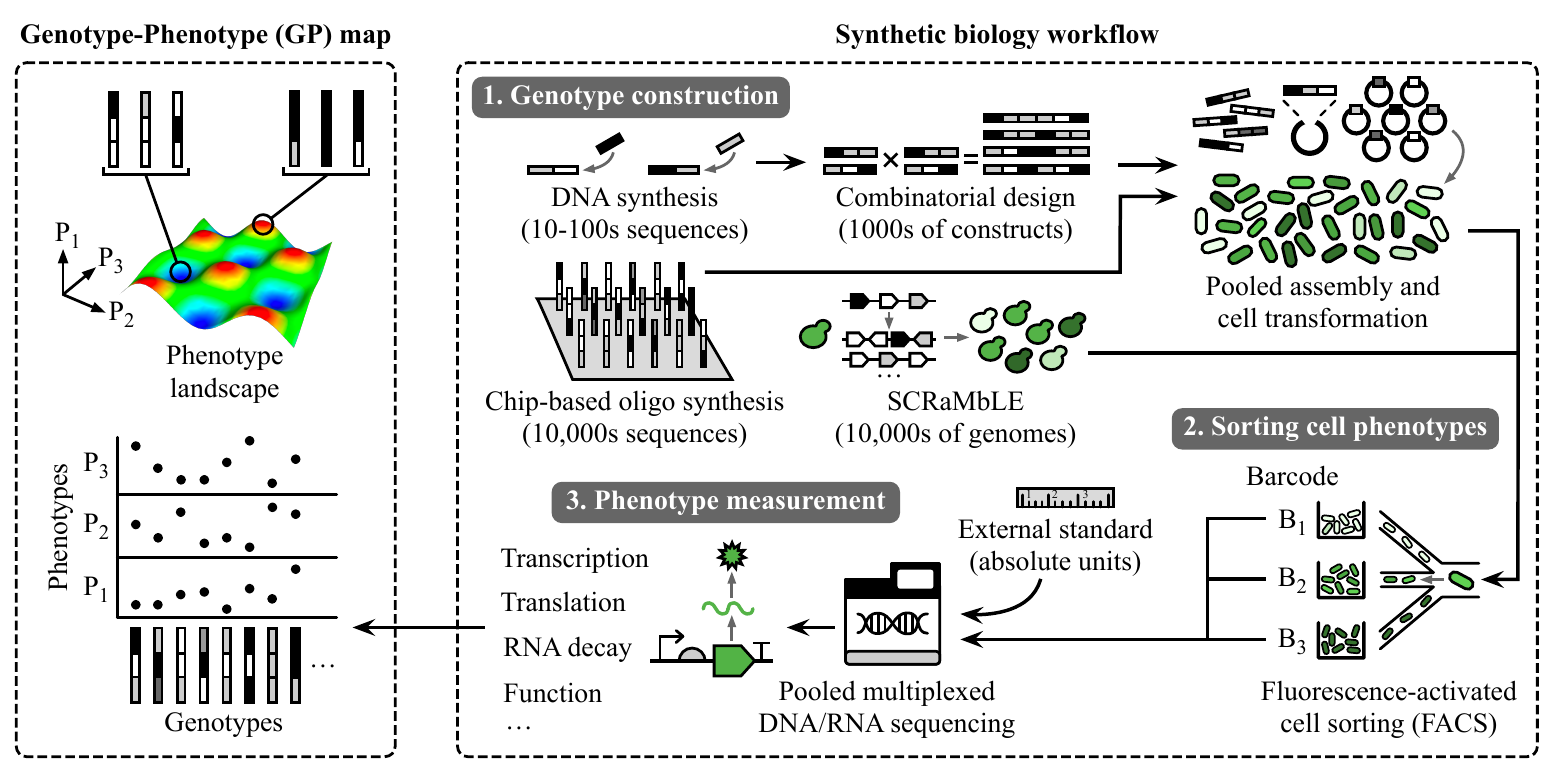}
\caption{
Synthetic biology methodologies can support the construction of detailed GP maps. Diverse sets of genotypes in cells can be generated using combinatorial DNA assembly \cite{woodruff:2017,gorochowski:2014,plesa:2018}, chip-based DNA synthesis \cite{kosuri:2014}, or systems to induce structural DNA rearrangements, e.g. Synthetic Chromosome Recombination and Modification by LoxP-mediated Evolution (SCRaMbLE) \cite{blount:2018}. Pooled libraries of cells can then be sorted into physically separated groups based on a parameter of interest, e.g. fluorescence of cells using fluorescence-activated cell sorting (FACS). Barcoded sequencing libraries can be generated from cells in each group and deep-sequencing of DNA/RNA performed to measure a wide range of phenotypic properties \cite{cambray:2018,gorochowski:2019,gorochowski:2017,kosuri:2013,johns:2018,gasperini:2016,liszczak:2018}. The inclusion of external standards during the sequencing allows for the conversion of relative phenotypic measures into absolutes units that are comparable across contexts \cite{gorochowski:2018}. Genetic diagrams drawn using Synthetic Biology Open Language Visual notation \cite{der:2017}. Image courtesy of Thomas E. Gorochowski.
}
\label{fig:paps}
\end{figure*}

A major goal in the field of synthetic biology is the re-purposing of biological components and systems to create living cells with new, designed functionalities. So what is the link between synthetic biology and GP maps? Faced with the challenge of understanding the function of biological parts and using this insight to rationally engineer cells, synthetic biologists frequently assemble large numbers of genetic designs (genotypes) and measure key aspects of the resultant cellular phenotypes. In doing so, novel methods for characterising GP maps have been developed (Fig.~\ref{fig:paps}). Key to many of these are two capabilities. First, it is necessary to be able to construct large numbers of diverse genotypes (referred to as {\it libraries}) in a structured way. For example, assembling many genetic circuits simultaneously, each one containing a different combination of functional DNA parts (e.g. protein coding genes or regulatory elements like promoters, ribosome binding sites and terminators) \cite{brophy:2014,gasperini:2016,appleton:2017}. Second, it should be possible to test these designs \textit{en masse}. To support both requirements, high-throughput, pooled DNA assembly and sequencing methods have been developed to measure the phenotype of every genetic circuit design (genotype) across huge libraries, effectively creating a detailed GP map.

Genotype libraries can be constructed in many ways, each with their own advantages and pitfalls. Perhaps the simplest and most transferable protocol involves pooled synthesis of a large library of pre-defined DNA parts, insertion of each part into a circular plasmid backbone that enables self-replication in cells, and transformation of the resulting plasmid library in the host cell of interest. This method can be used in combination with oligo(nucleotide) library synthesis (OLS) \cite{kosuri:2014} for generation of the DNA part library. Whilst limitations include genotype length (up to 200 nucleotides) and accuracy (error rate of 1 in 200 nucleotides), accessible genotype libraries allow access to regions of genotype space distant from one another, with the latest OLS derived study characterising 244,000 sequences simultaneously \cite{cambray:2018}. Other approaches for constructing libraries of genotypes include multiplexed DNA assembly \cite{plesa:2018,woodruff:2017,hughes:2017}, and site-specific incorporation of random genetic diversity \cite{komura:2018,patwardhan:2009,cozens:2018,holmqvist:2013}. The latter approach was recently used to characterise millions of promoter variants \cite{deboer:2020}.

Multiplexed measurement of many different phenotypes of the constructed genotype library is possible \cite{cambray:2018,gorochowski:2018}, though it must be ensured that each genotype contains a unique nucleotide-encoded barcode, to enable sequencing reads to be matched to the correct genotype \cite{church:1988}. Sequenceable phenotypes such as DNA or RNA abundance \cite{cambray:2018,johns:2018,kosuri:2013,patwardhan:2009} can be studied directly, in absolute units \cite{gorochowski:2019,gorochowski:2017}. Non-sequenceable phenotypes can be measured too, by sorting phenotypes into groups and then appending a unique barcode sequence to genotypes in each group (Fig.~\ref{fig:paps}). In this way, genotypes are mapped to phenotype categories. A detailed framework for design of pooled sequencing experiments (Multiplexed Assays for Variant Effects, MAVEs or Massively Parallel Reporter Assays, MPRAs) is available \cite{gasperini:2016}.

Much of the focus to date has been on using these methods to characterise genetic part function, measuring the behaviour of parts taken from distantly related species or designed \textit{in silico}. However characterisation of mutationally-connected genotype networks elucidates structural properties of GP maps for phenotypes which have not previously been characterised empirically at such scale. The resulting data can enable the construction of new \textit{in silico} models for predicting phenotypes from genotypes \cite{payne:2018,cuperus:2017} (section~\ref{sec:EEM}), a goal common to both synthetic biology and GP map researchers. Indeed, the synergy goes both ways: GP map studies highlight important principles which are only beginning to be considered by synthetic biologists during genetic circuit design. A clear example is genotypic robustness to mutations \cite{payne:2018}, which may prove important for genetic circuit longevity \cite{bull:2017,sleight:2013}. New science and technology brings new questions: ecological implications of microorganisms containing mutationally robust synthetic sequences have yet to be considered.

Expansion of this approach to study GP maps for different genotypes and phenotypes lies ahead. Innovations in nucleic acid sequencing are beginning to open up high-throughput characterisation of new types of phenotype in detail without sorting, such as epigenetic signatures or protein concentrations \cite{yus:2017,sze:2017,liszczak:2018}. The advent of long read sequencing \cite{dijk:2018} beckons high-throughput characterisation of GP maps for whole-cell genotypes. This is becoming possible with methods for high-throughput genome modification \cite{chari:2017}, such as SCRaMbLE which uses recombination for \textit{in vivo} combinatorial genomic rearrangement \cite{blount:2018}.

Pooled DNA assembly and sequencing is by no means the final solution for synthetic biologists or GP map researchers: crucially, it is limited by the number and length of assembleable genotypes and to phenotypes that can be inferred from sequencing data or for which high-throughput sorting methods (e.g.~FACS) exist. Nonetheless, this approach offers a significant increase in the size of GP maps that can be studied empirically and highlights the potential for mutually beneficial collaboration across these two emerging areas of biological research.

\begin{figure}[ht]
\begin{center}
\includegraphics[width=8.5cm]{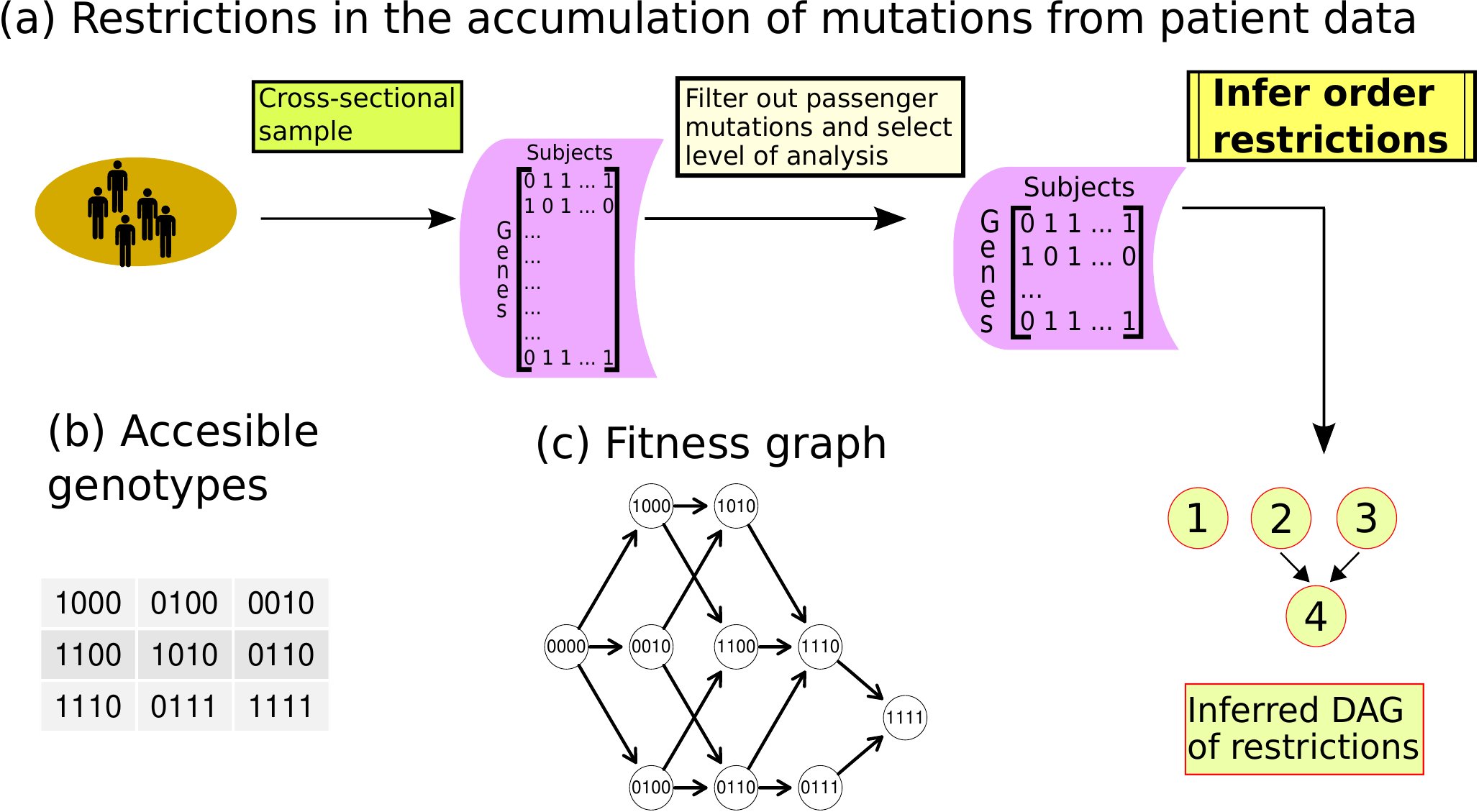}
\end{center}
\caption{Cancer progression models. (a) Main steps in the analysis of patient data. On the right, the DAG of restrictions shows genes in the nodes; an arrow from gene $i$ to gene $j$ indicates that a mutation in gene $i$ must occur before a mutation in gene $j$ can occur and, thus, indicates a direct dependency of a mutation in gene $j$ on a mutation in gene $i$. The absence of an arrow between two genes means that there are no direct dependencies between the two genes.  According to this DAG a mutation in the fourth gene can only be observed if both the second and third genes are mutated, but mutations in the first, second, and third gene do not have any dependencies among themselves. (b) Genotypes that fulfil the restrictions encoded in the DAG of restrictions: these are the accessible genotypes under the DAG. Genotypes are shown as sequences of 0s and 1s, where ``1100'' means a genotype with the first and second genes mutated. (c) Fitness graph or graph of mutational paths between accessible genotypes; nodes are genotypes (not genes) and arrows point toward mutational neighbors of higher fitness (thus, two genotypes connected by an arrow differ in one mutation that increases fitness \cite{crona:2013,devisser:2014}). Under CPMs, each new driver mutation with its dependencies satisfied increases fitness; therefore, all accessible genotypes that differ by exactly one mutation are connected in the fitness graph and the genotype with all driver genes mutated is the single fitness maximum. The fitness graph shows all the paths of tumor progression that start from the ``0000'' genotype and end in the fitness maximum.  Figures modified from \cite{diaz-uriarte:2019,diaz-uriarte:2015}.} 
\label{fig:cpm}
\end{figure}

\section{Consequences of GP maps for models of tumour evolution and cancer progression}
\label{sec:cancer}

Epistatic interactions between genetic alterations can constrain the order of accumulation of mutations during cancer progression (e.g. in colorectal cancer, mutations in the \emph{APC} gene are an early event that generally precedes mutations in the \emph{KRAS} gene \cite{gerstung:2011}). Cancer progression models (CPMs) have been developed to try to identify these restrictions during tumour progression using cross-sectional mutation data \cite{beerenwinkel:2015,beerenwinkel:2016}. CPMs take as input a cross-sectional sample from a population of cancer patients: each individual or patient provides a single observation, the cancer genotype in that patient. Thus, the input for CPMs is a matrix of individuals or patients by alteration events, where each entry in the matrix is binary coded as mutated/not-mutated or altered/not-altered (Fig.~\ref{fig:cpm}). The output from CPMs are directed acyclic graphs (DAGs) that encode the restrictions inferred (which are in fact sign epistasis relationships \cite{crona:2013,diaz-uriarte:2018}). In these DAGs, an edge between nodes $i$ and $j$ is to be interpreted as a direct dependence of an alteration of event $j$ on an alteration of event $i$; $j$ should never be observed altered unless $i$ is also altered. CPMs regard different patients as replicate evolutionary experiments, assume that the cancer cells in all patients are under the same genetic constraints \cite{beerenwinkel:2016,beerenwinkel:2015,gerstung:2011}, and ignore back mutations in the alteration of driver events. Thus, CPMs implicity encode all the possible mutational paths or trajectories of tumour progression (Fig.~\ref{fig:cpm}) \cite{diaz-uriarte:2019}, and some methods (e.g., CBN) provide estimates of the probabilities of the different paths of tumour progression \cite{diaz-uriarte:2019,hosseini:2019}. As in other domains, such as predicting antibiotic resistance, even small increases in our capacity to predict disease progression would be valuable for diagnostic, prognostic, and treatment purposes \cite{toprak:2012}; this renders CPMs a potentially useful tool in precision medicine. (Note that the focus here is on predicting mutational paths, but see also Section \ref{sec:empiricaltest} for transition forecasts, a different objective and approach when predicting evolution using GP maps.)

Several CPM methods have been developed, including oncogenetic trees (OT) \cite{szabo:2008,desper:1999}, conjunctive Bayesian networks (CBN) \cite{gerstung:2009,gerstung:2011,montazeri:2016}, and CAncer PRogression Inference (CAPRI) \cite{ramazzotti:2015,caravagna:2016}. The different methods differ in their model fitting procedures and in the types of restrictions they can represent. For example, OT can only return trees, where a mutation in a given gene has a direct dependence on only one other gene mutation; this is in contrast to CAPRI and CBN, where a mutation in a gene can depend on mutations in two or more different genes, and thus CAPRI and CBN return as output DAGs where some nodes can have multiple parents, as shown in Fig.~\ref{fig:cpm}. All CPMs focus on ``driver alterations'', i.e. those believed to actually drive, through selection, cancer progression (in contrast to so-called passenger mutations or hitchhikers). The types of alterations studied in CPMs range from changes in genes and pathways to gains and losses of chromosomal regions \cite{gerstung:2011,caravagna:2016,desper:1999}.

CPMs model sign epistasis, but they cannot model reciprocal sign epistasis (see section~\ref{box:epistasis}) \cite{diaz-uriarte:2018}, and thus CPMs effectively consider fitness landscapes with a single global peak. As a consequence, predictions of tumour progression, compared to the true paths of tumour progression, are very poor under multi-peaked fitness landscapes \cite{diaz-uriarte:2019}. Remarkably, even in the latter scenario, CPMs could be used to estimate an upper bound to the true evolutionary unpredictability \cite{diaz-uriarte:2019}; the analysis of twenty-two cancer data sets shows many of them to have low unpredictability.

CPMs do not force us to try to infer restrictions in the order of accumulation of alterations at any particular level or layer, in so far as the alterations examined can be regarded as heritable alterations with no back mutations. Thus, we could use the layers or levels of analysis (see also section \ref{sec:multilevel}) that are more relevant (e.g., metabolic pathways) or more likely to satisfy assumptions. Germane to this task are phenotypic bias (see section \ref{sec:RNABias}) and the effect on evolvability of many-to-many GP maps, relevant in the context of cancer \cite{nichol:2019,frank:2012}. These results suggest examining which is the most appropriate layer of analysis when using CPMs, which might not be gene alterations, but a layer closer to a ``heritable phenotype''. On the one hand, layers other than genes could allow us to maximise predictive ability (related to ideas on how to choose the relevant phenotypic dimensions \cite{altenberg:2005}). On the other hand, at other layers of analysis CPMs' assumptions might be more likely to be satisfied---in particular, the lack of reciprocal sign epistasis and local fitness maxima, as well as the absence of disjunctive (OR) relationships in dependencies between alterations (when a mutation in a gene can happen if a mutation in at least one of its parents has occurred; in Fig.~\ref{fig:cpm}, under an OR model, a mutation in gene 4 would need one of genes 2 or 3 to be mutated, but not both) \cite{diaz-uriarte:2018,diaz-uriarte:2019}.

CPMs assume Markovian evolution. However, non-Markovian dynamics on neutral networks \cite{manrubia:2015} (see also section \ref{sec:dynamics}) raises issues about choosing the layer of analysis for CPMs. For example, it seems unlikely that we could detect the existence of non-Markovian evolution reliably from the cross-sectional data used by CPMs. Additionally, non-Markovian evolution might be strong enough to cancel out possible benefits of working at other layers, and it could even be having an effect at the usual gene level of analysis where we label genes as altered/not-altered (mutated/not-mutated), because there is a many-to-one mapping between mutations in individual DNA bases and ``altered'' gene status.

The effects that environmental changes might have on evolutionary dynamics, given the dependence of epistatic relationships and fitness landscapes on the environment \cite{nichol:2019,lalic:2013,cervera:2016:JV,yubero:2017,payne:2019} (see also sections~\ref{sec:evolutionOFgpmaps} and \ref{sec:viruslandscape}), could be particularly relevant for the use of CPMs if, as posited by the ``adaptive oncogenesis'' hypothesis \cite{degregori:2018}, a key contribution to the relationship between age and cancer is the change in tissue fitness landscape with age (briefly, under the fitness landscapes of youth most mutants would have low fitness, unlike in the landscapes at older age). At a minimum, stratification of data sets by age would be warranted.

The sheer size of genotype and phenotype spaces is a potential matter of concern for CPMs, since the latter can only analyse a limited number of events. High-dimensional fitness landscapes might show increased mutational accessibility \cite{gavrilets:1997} and thus show both increased evolvability \cite{payne:2019} and decreased evolutionary predictability. From the point of view of predicting tumour evolution with CPMs, robustness to alterations in the features examined by the CPM would of course be a hurdle; but then, hopefully, these features would not have been regarded as ``drivers''. However, it should be mentioned here that ``passenger'' mutations in cancer, traditionally considered neutral, might actually reduce fitness of cancer cells and prevent tumour progression \cite{mcfarland:2017}; this raises the question of how to incorporate this lack of robustness in CPMs and, more generally, the extent of robustness and fitness landscape navigability in the cancer genome. The existence of a large pool of mildly deleterious passengers can also have consequences for procedures, such as CPMs, that analyse only a small subspace of the GP map. 

Of note, CPMs are often used in cancer progression scenarios where aneuploidies and karyotipic changes are common \cite{heng:2017}. This becomes an example of evolution of the GP map \cite{cuypers:2012,cuypers:2014,cuypers:2017}, a question deeply related to the proper comparison between GP maps of variable sequence length (see section~\ref{sec:evolutionOFgpmaps}). Choosing the right layer of analysis might again alleviate this problem, at least from the point of view of using CPMs to predict tumour evolution. Possible applicability of concepts emerging from evolving GP maps to the cancer genome are intriguing, especially given the possible costs of chromosomal instability and aneuploidy in cancer \cite{mcfarland:2017}, with the caveat that cancer constitutes a short-term evolution experiment that starts from cells with a long evolutionary history and that dies with its host \cite{kokko:2017}. Finally, the extent to which neutrality and phenotypic bias (see sections~\ref{sec:RNABias} and \ref{sec:dynamics}) affect CPMs remain as open questions, since CPMs are predicated on the idea that natural selection is what matters for the features studied.

\section{Summary and short-term perspectives}
\label{sec:perspectives}

Exhaustive enumerations of genotype spaces are only feasible for short sequence lengths. These enumerations may be sufficient in specific empirical cases, as to study transcription factor binding sites (see Sections~\ref{sec:SMevol} and \ref{sec:EEM}) or to build, in the near future, the first complete RNA GP maps incorporating experimentally measured fitness (SELEX experiments of small synthetic aptamers exploring the whole sequence space of length 24, such as those of Ref. \cite{jimenez:2013}, are already available). However, the number of possible genotypes for most biologically relevant sequence lengths is out of reach and, in the vast majority of cases, will always be: The estimated number of particles in the universe is of order $10^{80}$, a quantity comparable to the number of RNA sequences of length $L=133$ (the shortest known viroid has length $L=246$) or to that of proteins with $62$ amino acids (the class of ``small proteins'' refers to those with fewer than 100 amino acids). 

On the other hand, complete GP maps using RNA folding, the HP model, and toyLIFE \cite{garcia-martin:2018}, or using transcription factor binding \cite{berg:2004,khatri:2015a}, have proven to be very valuable resources for unveiling and testing some general properties of GP relationships, which seem to be common to several models. Further efforts toward theoretical developments that allow extrapolations to arbitrarily large genotype sizes, as well as approaches targeting higher levels of abstraction to study GP relationships without exploring the whole genotype space \cite{garcia-martin:2016b}, appear as two main avenues to complement computational studies. Though the specifics of folding algorithms do not seem to affect the statistical properties of GP maps, we cannot forget that the predictive abilities of those algorithms depend on the accuracy of the energy model and its parameters, which in the case of RNA or proteins are extrapolated from experimental measurements obtained under very specific conditions. Therefore, any improvement on this aspect will have a huge impact, not only in the accuracy of RNA, proteins, and possibly other GP maps, but in every related research field concerned with functional prediction. Computational analyses might also benefit from approaches that do not demand an exhaustive enumeration, but are tailored to test theoretical predictions, for example. One of them might be computations of the dual partition function of multiple RNA structures. Also, complete inverse folding methodologies can be used to develop computational frameworks for the study of genotype-phenotype-function relationships. Current algorithms can potentially build partial GP maps focused on phenotypes of interest, in which experimental data available can be fit, thus providing an appropriate context to make predictions and guide further experiments. New tools able to produce reliable estimates of structural properties, like neutral set size, robustness or evolvability, should be ideally independent of the GP map, as well as experimentally compatible, i.e.~they should allow predictions from small samples of genotypes. In this context, a sampling method that produces a genotype sample that optimally represents the phenotype of interest would be an important advance. There was some progress towards this goal in the form of a computational tool \cite{jorg:2008} which produces estimates of the size and robustness of RNA secondary structure phenotypes---but is in principle transferable to other GP maps---or through the estimation of the versatility of genotypes (see Section~\ref{sec:ConstrainedPositions}), but ample space for improvement remains. Also, and while these tools can predict neutral set size and robustness, no approaches exist yet for estimating phenotype evolvability or phenotype-phenotype correlations, which would yield the framework required to understand the evolution of evolvability or the deeply related concept of selection of the mutational neighbourhood. 

We are only beginning to understand how the structure of GP maps depends upon environmental conditions \cite{devos:2015,steinberg:2016,li:2018,gorter:2018}. We mostly ignore how the structure of a GP map changes with the dimensionality of genotype space, a topic that, beyond simple evolvable cells \cite{cuypers:2017} or toyLIFE \cite{arias:2014}, could potentially be explored using artificial genetic codes \cite{zhang:2017,fredens:2019} or expanded nucleotide alphabets \cite{hoshika:2019}. Finally, no matter the technological advance, the hyper-astronomical size of genotype space precludes the experimental construction of exhaustively-enumerated GP maps for large macromolecules, gene regulatory circuits, and metabolic pathways \cite{louis:2016}. This inconvenient fact necessitates the development of methods that can reliably infer the structure of a GP map from a relatively small sample of the map \cite{otwinowski:2014,duplessis:2016}. Besides analytical approaches based on generic properties of GP maps that allow inferences of their large-scale structure (see section~\ref{sec:UnivTopology}), advances in deep learning are already offering promising solutions to this key problem \cite{riesselman:2018}.

\subsection{Towards an improved understanding of GP map architecture: Is it universal?}

As discussed in section~\ref{sec:UnivTopology}, notable similarities exist amongst GP map properties giving rise to the notion of ``universal'' \cite{ahnert:2017,greenbury:2014} properties of GP maps, such as genetic correlations and phenotypic bias. Phenotypic bias, genetic correlations and evolvability are discussed in most studies of GP maps, but other properties, such as the assortativity of neutral networks \cite{aguirre:2011}, have only been analysed for some models. These topological properties could either provide a way of distinguishing between sequence-to-structure and artificial life GP maps or they could also be ``universal'' across a variety of models. At present, the universality of the structure induced in genotype spaces by evolutionarily sensible GP maps is a conjecture that those analyses, among others, could help to prove or disprove. Behind this conjecture, there is the main question of which fundamental mechanisms are responsible for the potentially universal features. As discussed in Section~\ref{sec:PossibleRoots}, spaces of high dimensionality that facilitate interconnections between genotypes and phenotypes seem to be a must. 

More specific explanations for the striking similarities detected among dissimilar GP maps have come from simple analytic models \cite{greenbury:2015,manrubia:2017} (see section~\ref{sec:ConstrainedPositions}). These models differ, but qualitatively they are all based on the fact that, depending on the phenotype, a part of the genotype is more constrained than the rest, for example to enable base pairing in RNA \cite{manrubia:2017}. Interestingly, such a model can also be constructed to predict GP map properties in Richard Dawkins' biomorphs \cite{martin:2020}. The fact that such widely different GP maps can be understood with similar models, supports the hypothesis that sequence constraints are an important cause for the observed similarities between GP maps. A question for future research is the extent to which these sequence constraints generalise to other biological or artificial life GP maps. Are there any counter-examples? And do the kind of assumptions about sequence constraints in the analytic models always hold or can we observe GP maps with similar properties which cannot be modelled based on sequence constraints? 

In the context of mathematical models of GP maps, it would also be desirable to further develop the existing models to explain more complex and biologically relevant situations, and to find out whether generic structural properties of genotype spaces are maintained under those circumstances. More realistic models should include mutations other than point mutations, such as deletions, duplications or insertions (there are just a few examples where the genome size is variable, among them that described in Section~\ref{sec:VirtualCells}), and recombination (see Section~\ref{sec:Recombination}). Extension to many-to-many GP maps by allowing multiple and semi-optimal phenotypes for a genotype---as it is the case for RNA sequences, for which there can exist multiple secondary structures with quite similar free energies---seems essential to fully understand adaptability \cite{deboer:2012,deboer:2014}. Models such as toyLIFE and virtual cells should be further studied if we want to explore issues relevant to synthetic biology, among others. A very relevant question for the synthetic biology community has been to design gene regulatory circuits that are mutationally robust \cite{chen:2011}. Results with toyLIFE show that genotypic robustness is a function not only of the individual components, but also of the complete network, which could be designed to be robust even if the individual components are not \cite{catalan:2018}. Actually, these extended models would have to come along with redefinitions of structural properties like neutral set size, robustness and evolvability. 

\subsection{Evolution \emph{on} and \emph{of} genotype spaces}

Since the eventual aim of GP map studies is to understand evolutionary processes, a key question is how each of the universal GP map properties---should they exist---affect evolutionary outcomes. Phenotypic bias, for example, implies that only very abundant phenotypes will be visited when adapting to a new evolutionary challenge. This implies that the evolutionary search is constrained to look in the space of very abundant solutions. This constraint might lead to a limitation in the number of possible phenotypes attainable through evolution: it has been put forward, and supported with simple developmental models, that the small fraction of phenotypes visible to evolution are highly clustered in morphospace and that the most frequent phenotypes are the most similar \cite{borenstein:2008}---recalling the relevance of phenotype-phenotype correlations. Since evolutionary search is a consequence of the stochastic nature of the evolutionary dynamics, and is not dependent on the particulars of the GP map, there is no reason why this phenomenon should not be observed in real GP maps. 

It has been also shown that transition times between phenotypes depend very strongly on how they are connected in genotype space, and there is also a strong indication that the genotypic robustness in a neutral network plays a role \cite{hu:2012}: transition times between phenotypes depend strongly on how accessible a given phenotype is from the most robust genotypes. Because evolution naturally tends to visit the most robust genotypes \cite{nimwegen:1999}, their connections to other phenotypes may be more relevant than those of less robust genotypes. The natural question to address is if there is a mutational bias in evolution towards phenotypes that connect to robust genotypes. Though computational GP maps have been the primary tool to explore this question in depth, some experimental work on this topic has been carried out as well. Indeed, {\it Pseudomonas aeruginosa} preferentially chooses three particular mutational pathways to evolve an adaptive phenotype under certain conditions \cite{lind:2015}. When these pathways are repressed (through gene knockouts), the bacteria are able to evolve the same phenotype, but using new mutations. Actually, those mutations were available in the original population, but the probability of fixing them is very small compared to the three preferred pathways. This work gives empirical support to the relevance of mutational neighbourhoods for evolution (see Section~\ref{sec:Quasispecies}), and highlights once more the need for further computational and experimental investigations of this topic. 

Our knowledge of how the GP map properties individually affect evolutionary outcomes is still incomplete. For example, phenotypic bias is known to affect evolutionary outcomes due to at least three mechanisms: the `survival of the flattest' \cite{wilke:2001Nat}, the `arrival of the frequent' \cite{schaper:2014}, and its effect on the free fitness of phenotypes in monomorphic regime \cite{iwasa:1988,sella:2005,khatri:2009,khatri:2015a}. Despite this progress, it may still be difficult to estimate for more complex cases than the scenario studied by Schaper and Louis, how strong the `arrival of the frequent' effect will be and whether phenotypic frequency or phenotypic fitness are likely to determine evolutionary outcomes. Ultimately, this knowledge will help us answer the bigger questions of whether and how we can use GP maps to predict short- and long-term trends in evolution \cite{lassig:2017,milocco:2019,nosil:2020}.

The application of the tools of network science to the previous context, and to evolutionary dynamics of heterogeneous populations at large, opens a promising avenue that, as of today, faces however some limitations and difficulties. First, and although the hyperastronomical sizes of genotype networks seem an insurmountable obstacle, the theory of competing networks shows that, in genotype spaces where function is relatively sparse, only the much smaller local subnetworks are relevant to analyse the evolution of populations---while the rest of the huge network of networks is in practice negligible \cite{yubero:2017}. A different promising avenue is the generic construction of a phenotype network that can be computationally---and likely analytically---tackled \cite{cowperthwaite:2008,schaper:2014,manrubia:2015}. Second, most theoretical work has been developed using models that only consider point mutations. The introduction of different mutational mechanisms, as discussed in the previous sections, would drastically transform the topology and spectral properties of genotype networks. However, once the new network defined through those rules is known, the analysis proceeds following standard procedures. When the GP map is many-to-many, either due to environmental changes or to phenotypic promiscuity, more complex configurations such as multi-layer networks should be introduced to properly describe the evolution of the system \cite{catalan:2017,aguirre:2018}. Recombination cannot be easily cast in this network framework, which is unable to describe the process in detail \cite{azevedo:2006,devisser:2009,paixao:2014}. Different approaches can be however used in this case (see Section~\ref{sec:Recombination}) and hopefully combined to eventually yield a unified formal description of dynamics under a variety of microscopic processes generating diversity.

\section{Outlook: On the feasibility of a complete genotype-to-organism map}
\label{sec:GOmap}

Systems such as RNA folding, protein secondary structure, and transcription factor binding are attractive models for understanding the GP map because it is possible to compute the map from physical first-principles. But these processes are only the first steps in the long chain of interactions whose end result is organismal function, structure, viability, and reproduction \cite{bershtein:2017}. At these higher levels of integration, it is the integration itself that in large measure determines the GP map. To address the question of whether evolution \emph{of} the GP map or evolution \emph{on} the GP map is the appropriate framework, it may be helpful to use a distinction \cite{altenberg:1995} of two different properties of the GP map: \emph{generative} properties---how the genotype is actually used to produce the phenotype---and \emph{variational} properties---the way that changes in the genotype map to changes in the phenotype.  More recently, this distinction has been called ``formative'' and ``differential'' properties, respectively \cite{orgogozo:2015}.

Unravelling the generative properties of the GP map is the main agenda of molecular and developmental biology. The variational properties ultimately derive from the generative properties, so the question is whether anything systematic about either can be predicted from evolutionary theory. Tremendous resources have been dedicated to molecular and cellular biology with the promise that, by identifying all the parts and interactions involved in a biological phenomenon, it could be understood, controlled, and even synthesised. The fruition of this promise has been realised in many cases, as attested to by the advent of successful treatments for many diseases. This was the justification of the Human Genome Project, with the hope that once all of the human DNA sequence was known, the genetic basis of diseases and organismal functions would be attainable. But a surprise from the Human Genome Project was the ``missing heritability'' that emerged from genome-wide association studies (GWAS). Analysis of DNA sequences could identify only a small fraction of the genes responsible for human phenotypes known from family studies to have high heritability.

Currently there are contradictory findings about the GP map at the whole organism level. On the one hand are studies which find that organisms exhibit a modular structure over large classes of phenotypic variables \cite{wagner:2011b}. to the point where modularity is often stated as an accepted fact \cite{espinosa-soto:2018}. On the other hand are studies which find that almost every gene affects many characters (universal pleiotropy) and almost every character is affected by many genes, summarised as the \emph{omnigenic model} of the GP map \cite{boyle:2017}. The omnigenic model proposes that while there may be ``core genes'' contributing to any given phenotype, the network of gene-interactions has a ``small world'' topology, a property that leads to broad pleiotropy and polygeny in the GP map. 

It may prove helpful that there is another field that is also trying to understand how complex functional behaviours emerge out of the interaction of thousands or millions of simple parts---the artificial neural network community. Artificial neural networks (ANNs) have now been created whose behaviours rival or exceed certain human cognitive capabilities. Because of the recent achievements of ANNs, the field is currently in an explosive state of development. The achievement of the engineers outpaced the understanding of the theoreticians as to why deep learning networks perform so well, and the theoreticians are working to catch up. As of now, there are several observations about ANNs that may be instructive to those making computational models of the GP map.

The multilayered ``deep neural networks'' (DNNs), which have proven to be the most successful ANNs, are defined by a collection of thousands or millions of algorithmically learned numbers. The numbers specify the weights of connections between nodes, and each node sums its inputs from other nodes, and then outputs a function of this sum as inputs to other nodes. What is most notable about DNN engineering is that there is very little interest in the specific values of the numbers, and no way to understand how the specific numbers generate the network behaviour. While there has been some success at \emph{interpreting} DNNs---where one identifies what feature of an input causes a particular neuron in the network to activate---there is currently little understanding of how the all weights connecting the neurons produce this behaviour. The main focus has been on the \emph{processes} that generate the numbers, and this is where theoreticians are attempting to generate understanding. The most successful process for training the weights is based on their variational properties: how changes in the weights change the error between the network's actual behaviour and its desired behaviour. The methods of back-propagation and stochastic gradient descent change the weights until there is little or no error on a set of traning examples applied to the network \cite{bottou:2010}. The variational properties of greatest interest are how the network behaves on novel inputs, and how changes in the inputs map to changes in the network behaviour.

A similar situation may hold in complex organisms. Without understanding or even knowing how the thousands of organismal components are generating phenotypes on the whole-organism level, we may nevertheless be able to understand its variational properties based on evolutionary processes. Here we briefly list a few principal processes that are understood to shape the variational properties of genotypes.

\subsection{The evolution of re-evolvability under varying selection} 

In a number of different GP map models, evolution under recurring variation in natural selection moves the genome to places in genotype space where fewer and fewer mutations are needed to re-evolve previous adaptations when the old environment returns. This has been observed for a model of two neutral networks \cite{draghi:2008} (also called the evolution of ``genetic potential'' \cite{ancel:2005}), networks of logic gates (the variation is called ``modularly varying goals'' \cite{kashtan:2005}), gene regulatory networks and the virtual cells discussed above (just called the ``evolution of evolvability'' \cite{cuypers:2017,crombach:2007,crombach:2008}). However, not all GP maps support this phenomenon \cite{kashtan:2005}. Exactly what properties a GP map must possess to allow the evolution of re-evolvability remains an open problem. 

\subsection{Constructional selection} 

Genes not only provide material for the generation of the phenotype, but provide degrees of freedom for varying the phenotype. A gene duplication or {\it de novo} gene origin thus differs from a point mutation in that it increases the degrees of freedom of the GP map, and thus adds new variational properties to the genome. Gene duplications and deletions are frequent events in eukaryotic reproduction. Any variational property of a gene that is associated with the gene being retained in the genome can thus become enriched in the GP map \cite{altenberg:1995}. The likelihood of a duplicate copy of a gene to be retained by evolution has been called its ``gene duplicability'' \cite{yang:2003}. The identification of gene properties that are associated with gene duplicability is an active area of research. Some of the properties identified include:
\begin{itemize}
\item peripheral versus central position in protein-protein interaction networks \cite{chen:2014};
\item high levels of gene expression \cite{mattenberger:2017};
\item high rates of sequence evolution before duplication \cite{otoole:2017};
\item ordered, versus intrinsically disordered proteins \cite{banerjee:2017};
\item signaling, transport, and metabolism functions increase gene duplicability, while involvement in genome stability and organelle function reduces it, for whole genome duplications in plants \cite{liz:2016}.
\end{itemize}

While the causes of differential gene duplicability have been subject of a great deal of investigation, its \emph{consequences} for organismal evolvability have received limited thought. Quantitative models for how differences in gene duplicability can shape the variational properties of the entire genome\cite{altenberg:1995} have been applied to examples of evolutionary computation\cite{altenberg:1994:EEGP,altenberg:1994:EBR,altenberg:1994:EPIGP} under the rubric ``constructional selection''. One can conceive of the genome as a population of genes, and differences in gene duplicability as fitness differences, not on the organismal level, but on this level of genome-as-population.  Constructional selection results in the enrichment of the genome in genes that have a higher likelihood of being retained when copies of them are created. These are gene copies that evolve to where deletion or inactivation becomes deleterious to the organism. This occurs for genes more likely to subfunctionalise, or escape adaptive conflict, or neofunctionalise. It provides a ubiquitous mechanism for the evolution of evolvability.

\subsection{Entropic evolutionary forces}  

The GP map is mostly cast as a many-to-one map because there may be multiple genotypes that result in the same phenotype, due to low-level properties such as synonymous codons, but also due to multiple ways that ligand-receptor bonds may be achieved, and multiple ways that the same gene regulatory interactions may be encoded. This degeneracy of the GP map\cite{whitacre:2010} creates the possibility of evolution along neutral networks of mutationally-connected genotypes with the same fitness. The randomness of evolution along neutral networks brings forth statistical mechanical forces of entropy increase. This entropic behaviour has been described as ``biology's first law'' \cite{mcshea:2010}.

Entropic phenomena that result from evolution along neutral networks include:
\paragraph {\bf Subfunctionalization} If different functions in a gene are modular enough so that they can be individually disabled without affecting each other, then the process of gene duplication and complementary loss of functions effectively spreads the functions among multiple genes \cite{hughes:1994,stoltzfus:1999,force:1999}. Since there are many more ways to spread the functions apart than to keep them in one gene, there is an entropic force in the direction of separating separable functions.
\paragraph {\bf Constructive neutral evolution} Stoltzfus \cite{stoltzfus:1999} introduced the general concept of entropic processes that produce greater genetic complexity to traits simply because there happen to be more complex ways to generate a trait than there are simple ways. If there are neutral mutational pathways between alternate means of generating traits, then the more numerous class will come to dominate.
\paragraph{\bf Non-optimal phenotypes} As explored in Section~\ref{sec:SMevol}, in the context of the weak mutation monomorphic regime, there is an exact analogy to statistical mechanics, embodied in a quantity called free fitness \cite{iwasa:1988,khatri:2015}, which is the sum the fitness of phenotypes and the sequence entropy (log degeneracy) weighted by the analogue of temperature, the inverse of the population size. This means that for small populations, evolution gives rise to non-optimal phenotypes that balance fitness and entropy, or free fitness.
\paragraph {\bf Developmental systems drift} Primary sequences may diverge between species even while the same developmental outcomes are maintained \cite{true:2001}. Within the free fitness framework traits this has been explored (Section~\ref{sec:SMevol}) under stabilising selection, and for small populations it is predicted that the effect of sequence entropy is that populations develop isolation more quickly \cite{khatri:2019}.

The great variation in genome sizes over different taxa and even within closely related taxa suggests that the quantity of DNA maintained in the genome may function as a quantitative trait subject to species-specific natural selection. Just as in physical systems, where entropic forces can be counteracted by energy potentials, natural selection on genomic complexity as a quantitative trait may counteract the entropic tendencies in constructive neutral evolution.  Such dynamics may be at work in the genomic streamlining discussed in Section~\ref{sec:empirical}. As is seen with the infinitesimal model of quantitative genetics, even though any individual streamlining event or an individual complexification event may have unobservable effects on fitness, the aggregate forces of entropic complexification and quantitative selection on genome size may statistically push the genome toward a balance, the character of which depends on the species-specific costs of genome maintenance.

\subsection{Omnigenic integration} 

When adaptations are produced by large scale interactions of organismal components, the GP map can be expected to be highly polygenic. Complex interactions of many components make the individual components also highly pleiotropic. Any given genetic change may beneficially affect certain traits while being detrimental to others. When their net effect is beneficial, then they are selected, but the deleterious effects they produce on certain traits creates the opportunity for other genetic variation to compensate for these effects. The GP map then becomes a patchwork of compensatory effects. In the limit of small effects, this patchwork becomes Fisher's \emph{infinitesimal model,} \cite{barton:2017} in which pleiotropy and polygeny are continuous and ubiquitous and there is little structuring of the GP map.

\subsection{Selection for mixability}

Natural selection in sexual organisms with genetic recombination favors alleles that have high average fitness among all the different genotypes in which they appear in the population.  An allele which might produce a highly adaptive phenotype when combined with just the right alleles at the same or other loci faces the breakup of such an advantageous combination due to segregation and recombination.  Alleles which produce a reliable fitness advantage regardless of the genetic variation they are recombined with---a property called ``mixability''---have a selective advantage \cite{livnat:2008}.  The aggregate consequence of selection for mixability is toward greater modularity in the production of phenotypes:  alleles individually produce the adaptive advantage without reliance on particular states of alleles at other loci.  It suggests a process that counteracts the ``omnigenic'' model of complete genomic integration.  The consequences of selection for mixability on the GP map have only begun to be elucidated \cite{livnat:2010}.

\subsection{Epistatic smoothing of the fitness landscape}

Conrad noted that a mutation which smoothed the fitness landscape for other loci would enhance their chance of producing advantageous mutations, and hitchhike along with such mutations, thus providing a constant force toward reducing reciprocal sign epistasis. This is the earliest mechanism proposed for the evolution of evolvability \cite{conrad:1972,conrad:1979}, and has yet to be fully investigated theoretically.

\subsection{Summary}

We have identified a patchwork of processes that in principle are able to shape the variational properties of the GP map for phenotypes at the level of whole organisms, where complex integration leaves us unable to derive the properties from physical first-principles. This is an area in which evolutionary theory needs much greater development.  At levels of complexity at which detailed reductionist modelling is currently impossible, we have surveyed efforts to date that attempt to analyse how evolutionary processes shape the GP map. The body of results described, while not a fully fleshed-out theory, is perhaps sufficient to demonstrate that this process-based approach can inform a research program for the GP map at the whole organism level.

\section*{Acknowledgements}

All authors are indebted to the Centre Europ\'een de Calcul Atomique et Mol\'eculaire (CECAM) for supporting the organization of the workshop ``From genotypes to function. Challenges in the computation of realistic genotype-phenotype maps", which took place in Zaragoza (March 13th to March 15th, 2019) and triggered the production of this work.
These are additional sources of financial support of the authors: \\
SM: grant FIS2017-89773-P (MINECO/FEDER, EU); “Severo Ochoa” Centers of Excellence to CNB, SEV 2017-0712\\
JAC: grants FIS2015-64349-P (MINECO/FEDER, EU) and PGC2018-098186-B-I00 (MICINN/FEDER, EU) \\
JA: grant FIS2017-89773-P (MINECO/FEDER, EU) \\
LA: Foundational Questions Institute (FQXi) and Fetzer Franklin Fund, a donor advised fund of Silicon Valley Community Foundation, for FQXi Grant number FQXi-RFP-IPW-1913, Stanford Center for Computational, Evolutionary and Human Genomics and the Morrison Institute for Population and Resources Studies, Stanford University, the 2015 Information Processing in Cells and Tissues Conference, and the Mathematical Biosciences Institute at The Ohio State University, for its support through National Science Foundation Award \#DMS 0931642 \\
PC: Ram\'on Areces Postdoctoral Fellowship \\
RDU: grant BFU2015-67302-R (MINECO/FEDER, EU) \\
SFE: grants BFU2015-65037-P  (MCIU-FEDER) and PROMETEOII/2014/012 (Generalitat Valenciana) \\
JK: DFG within CRC1310 ``Predictability in Evolution'' \\
NSM: Gates Cambridge Scholarship; Winton Programme for the Physics of Sustainability \\
JLP: Swiss National Science Foundation, grant PP00P3\_170604 \\
MJT: grants EP/L016494/1 (EPSRC/BBSRC Centre for Doctoral Training in Synthetic Biology) and BB/L01386X/1 (BBSRC/EPSRC Synthetic Biology Research Centre, BrisSynBio) \\
MW: the EPSRC and the Gatsby Charitable Foundation






\bibliographystyle{elsarticle-num-names}
\bibliography{bibliography.bib}







\end{document}